\documentclass[a4paper,11pt]{article}
\pdfoutput=1 

\usepackage{jinstpub} 
\usepackage{mathrsfs}

\usepackage{subcaption}
\usepackage{siunitx}
\usepackage{color,soul}

\title{Highly-parallelized simulation of a pixelated LArTPC on a GPU}

\usepackage[symbol]{footmisc}

\collaboration{The DUNE Collaboration}
\affiliation[0]{Abilene Christian University, Abilene, TX 79601, USA}
\affiliation[1]{University of Albany, SUNY, Albany, NY 12222, USA}
\affiliation[2]{University of Amsterdam, NL-1098 XG Amsterdam, The Netherlands}
\affiliation[3]{Antalya Bilim University, 07190 D{\"o}{\c{s}}emealt{\i}/Antalya, Turkey}
\affiliation[4]{University of Antananarivo, Antananarivo 101, Madagascar}
\affiliation[5]{Universidad Antonio Nari{\~n}o, Bogot{\'a}, Colombia}
\affiliation[6]{Argonne National Laboratory, Argonne, IL 60439, USA}
\affiliation[7]{University of Arizona, Tucson, AZ 85721, USA}
\affiliation[8]{Universidad Nacional de Asunci{\'o}n, San Lorenzo, Paraguay}
\affiliation[9]{University of Athens, Zografou GR 157 84, Greece}
\affiliation[10]{Universidad del Atl{\'a}ntico, Barranquilla, Atl{\'a}ntico, Colombia}
\affiliation[11]{Augustana University, Sioux Falls, SD 57197, USA}
\affiliation[12]{University of Basel, CH-4056 Basel, Switzerland}
\affiliation[13]{University of Bern, CH-3012 Bern, Switzerland}
\affiliation[14]{Beykent University, Istanbul, Turkey}
\affiliation[15]{University of Birmingham, Birmingham B15 2TT, United Kingdom}
\affiliation[16]{Universit{\`a} del Bologna, 40127 Bologna, Italy}
\affiliation[17]{Boston University, Boston, MA 02215, USA}
\affiliation[18]{University of Bristol, Bristol BS8 1TL, United Kingdom}
\affiliation[19]{Brookhaven National Laboratory, Upton, NY 11973, USA}
\affiliation[20]{University of Bucharest, Bucharest, Romania}
\affiliation[21]{University of California Berkeley, Berkeley, CA 94720, USA}
\affiliation[22]{University of California Davis, Davis, CA 95616, USA}
\affiliation[23]{University of California Irvine, Irvine, CA 92697, USA}
\affiliation[24]{University of California Los Angeles, Los Angeles, CA 90095, USA}
\affiliation[25]{University of California Riverside, Riverside CA 92521, USA}
\affiliation[26]{University of California Santa Barbara, Santa Barbara, California 93106 USA}
\affiliation[27]{California Institute of Technology, Pasadena, CA 91125, USA}
\affiliation[28]{University of Cambridge, Cambridge CB3 0HE, United Kingdom}
\affiliation[29]{Universidade Estadual de Campinas, Campinas - SP, 13083-970, Brazil}
\affiliation[30]{Universit{\`a} di Catania, 2 - 95131 Catania, Italy}
\affiliation[31]{Universidad Cat{\'o}lica del Norte, Antofagasta, Chile}
\affiliation[32]{Centro Brasileiro de Pesquisas F\'isicas, Rio de Janeiro, RJ 22290-180, Brazil}
\affiliation[33]{IRFU, CEA, Universit{\'e} Paris-Saclay, F-91191 Gif-sur-Yvette, France}
\affiliation[34]{CERN, The European Organization for Nuclear Research, 1211 Meyrin, Switzerland}
\affiliation[35]{Institute of Particle and Nuclear Physics of the Faculty of Mathematics and Physics of the Charles University, 180 00 Prague 8, Czech Republic }
\affiliation[36]{University of Chicago, Chicago, IL 60637, USA}
\affiliation[37]{Chung-Ang University, Seoul 06974, South Korea}
\affiliation[38]{CIEMAT, Centro de Investigaciones Energ{\'e}ticas, Medioambientales y Tecnol{\'o}gicas, E-28040 Madrid, Spain}
\affiliation[39]{University of Cincinnati, Cincinnati, OH 45221, USA}
\affiliation[40]{Centro de Investigaci{\'o}n y de Estudios Avanzados del Instituto Polit{\'e}cnico Nacional (Cinvestav), Mexico City, Mexico}
\affiliation[41]{Universidad de Colima, Colima, Mexico}
\affiliation[42]{University of Colorado Boulder, Boulder, CO 80309, USA}
\affiliation[43]{Colorado State University, Fort Collins, CO 80523, USA}
\affiliation[44]{Columbia University, New York, NY 10027, USA}
\affiliation[45]{Comisi{\'o}n Nacional de Investigaci{\'o}n y Desarrollo Aeroespacial, Lima, Peru}
\affiliation[46]{Centro de Tecnologia da Informacao Renato Archer, Amarais - Campinas, SP - CEP 13069-901}
\affiliation[47]{Central University of South Bihar, Gaya, 824236, India }
\affiliation[48]{Institute of Physics, Czech Academy of Sciences, 182 00 Prague 8, Czech Republic}
\affiliation[49]{Czech Technical University, 115 19 Prague 1, Czech Republic}
\affiliation[50]{Dakota State University, Madison, SD 57042, USA}
\affiliation[51]{University of Dallas, Irving, TX 75062-4736, USA}
\affiliation[52]{Laboratoire d{\textquoteright}Annecy de Physique des Particules, Univ. Grenoble Alpes, Univ. Savoie Mont Blanc, CNRS, LAPP-IN2P3, 74000 Annecy, France}
\affiliation[53]{Daresbury Laboratory, Cheshire WA4 4AD, United Kingdom}
\affiliation[54]{Dordt University, 700 7th St NE, Sioux Center, IA 51250, USA}
\affiliation[55]{Drexel University, Philadelphia, PA 19104, USA}
\affiliation[56]{Duke University, Durham, NC 27708, USA}
\affiliation[57]{Durham University, Durham DH1 3LE, United Kingdom}
\affiliation[58]{University of Edinburgh, Edinburgh EH8 9YL, United Kingdom}
\affiliation[59]{Universidad EIA, Envigado, Antioquia, Colombia}
\affiliation[60]{E{\"o}tv{\"o}s Lor{\'a}nd University, 1053 Budapest, Hungary}
\affiliation[61]{ETH Zurich, Zurich, Switzerland}
\affiliation[62]{Faculdade de Ci{\^e}ncias da Universidade de Lisboa - FCUL, 1749-016 Lisboa, Portugal}
\affiliation[63]{Universidade Federal de Alfenas, Po{\c{c}}os de Caldas - MG, 37715-400, Brazil}
\affiliation[64]{Universidade Federal de Goias, Goiania, GO 74690-900, Brazil}
\affiliation[65]{Universidade Federal do ABC, Santo Andr{\'e} - SP, 09210-580, Brazil}
\affiliation[66]{Universidade Federal do Rio de Janeiro,  Rio de Janeiro - RJ, 21941-901, Brazil}
\affiliation[67]{Fermi National Accelerator Laboratory, Batavia, IL 60510, USA}
\affiliation[68]{University of Ferrara, Ferrara, Italy}
\affiliation[69]{University of Florida, Gainesville, FL 32611-8440, USA}
\affiliation[70]{Florida State University, Tallahassee, FL, USA}
\affiliation[71]{Fluminense Federal University, 9 Icara{\'\i} Niter{\'o}i - RJ, 24220-900, Brazil }
\affiliation[72]{Universit{\`a} degli Studi di Genova, Genova, Italy}
\affiliation[73]{Georgian Technical University, Tbilisi, Georgia}
\affiliation[74]{University of Granada {\&} CAFPE, 18002 Granada, Spain}
\affiliation[75]{Gran Sasso Science Institute, L'Aquila, Italy}
\affiliation[76]{Laboratori Nazionali del Gran Sasso, L'Aquila AQ, Italy}
\affiliation[77]{University Grenoble Alpes, CNRS, Grenoble INP, LPSC-IN2P3, 38000 Grenoble, France}
\affiliation[78]{Universidad de Guanajuato, Guanajuato, C.P. 37000, Mexico}
\affiliation[79]{Harish-Chandra Research Institute, Jhunsi, Allahabad 211 019, India}
\affiliation[80]{Harvard University, Cambridge, MA 02138, USA}
\affiliation[81]{University of Hawaii, Honolulu, HI 96822, USA}
\affiliation[82]{University of Houston, Houston, TX 77204, USA}
\affiliation[83]{University of  Hyderabad, Gachibowli, Hyderabad - 500 046, India}
\affiliation[84]{Idaho State University, Pocatello, ID 83209, USA}
\affiliation[85]{Institut de F{\'\i}sica d{\textquoteright}Altes Energies (IFAE){\textemdash}Barcelona Institute of Science and Technology (BIST), Barcelona, Spain}
\affiliation[86]{Instituto de F{\'\i}sica Corpuscular, CSIC and Universitat de Val{\`e}ncia, 46980 Paterna, Valencia, Spain}
\affiliation[87]{Instituto Galego de F{\'\i}sica de Altas Enerx{\'\i}as, University of Santiago de Compostela, 15782 Santiago de Compostela, Spain}
\affiliation[88]{Illinois Institute of Technology, Chicago, IL 60616, USA}
\affiliation[89]{Imperial College of Science Technology and Medicine, London SW7 2BZ, United Kingdom}
\affiliation[90]{Indian Institute of Technology Guwahati, Guwahati, 781 039, India}
\affiliation[91]{Indian Institute of Technology Hyderabad, Hyderabad, 502285, India}
\affiliation[92]{Indiana University, Bloomington, IN 47405, USA}
\affiliation[93]{Istituto Nazionale di Fisica Nucleare Sezione di Bologna, 40127 Bologna BO, Italy}
\affiliation[94]{Istituto Nazionale di Fisica Nucleare Sezione di Catania, I-95123 Catania, Italy}
\affiliation[95]{Istituto Nazionale di Fisica Nucleare Sezione di Ferrara, I-44122 Ferrara, Italy}
\affiliation[96]{Istituto Nazionale di Fisica Nucleare Laboratori Nazionali di Frascati, Frascati, Roma, Italy}
\affiliation[97]{Istituto Nazionale di Fisica Nucleare Sezione di Genova, 16146 Genova GE, Italy}
\affiliation[98]{Istituto Nazionale di Fisica Nucleare Sezione di Lecce, 73100 - Lecce, Italy}
\affiliation[99]{Istituto Nazionale di Fisica Nucleare Sezione di Milano Bicocca, 3 - I-20126 Milano, Italy}
\affiliation[100]{Istituto Nazionale di Fisica Nucleare Sezione di Milano, 20133 Milano, Italy}
\affiliation[101]{Istituto Nazionale di Fisica Nucleare Sezione di Napoli, I-80126 Napoli, Italy}
\affiliation[102]{Istituto Nazionale di Fisica Nucleare Sezione di Padova, 35131 Padova, Italy}
\affiliation[103]{Istituto Nazionale di Fisica Nucleare Sezione di Pavia,  I-27100 Pavia, Italy}
\affiliation[104]{Istituto Nazionale di Fisica Nucleare Laboratori Nazionali di Pisa, Pisa PI, Italy}
\affiliation[105]{Istituto Nazionale di Fisica Nucleare Sezione di Roma, 00185 Roma RM, Italy}
\affiliation[106]{Istituto Nazionale di Fisica Nucleare Laboratori Nazionali del Sud, 95123 Catania, Italy}
\affiliation[107]{Universidad Nacional de Ingenier{\'\i}a, Lima 25, Per{\'u}}
\affiliation[108]{Institute for Nuclear Research of the Russian Academy of Sciences, Moscow 117312, Russia}
\affiliation[109]{University of Insubria, Via Ravasi, 2, 21100 Varese VA, Italy}
\affiliation[110]{University of Iowa, Iowa City, IA 52242, USA}
\affiliation[111]{Iowa State University, Ames, Iowa 50011, USA}
\affiliation[112]{Institut de Physique des 2 Infinis de Lyon, 69622 Villeurbanne, France}
\affiliation[113]{Institute for Research in Fundamental Sciences, Tehran, Iran}
\affiliation[114]{Instituto Superior T{\'e}cnico - IST, Universidade de Lisboa, Portugal}
\affiliation[115]{Instituto Tecnol{\'o}gico de Aeron{\'a}utica, Sao Jose dos Campos, Brazil}
\affiliation[116]{Iwate University, Morioka, Iwate 020-8551, Japan}
\affiliation[117]{Jackson State University, Jackson, MS 39217, USA}
\affiliation[118]{University of Jammu, Jammu-180006, India}
\affiliation[119]{Jawaharlal Nehru University, New Delhi 110067, India}
\affiliation[120]{Jeonbuk National University, Jeonrabuk-do 54896, South Korea}
\affiliation[121]{Joint Institute for Nuclear Research, Dzhelepov Laboratory of Nuclear Problems 6 Joliot-Curie, Dubna, Moscow Region, 141980 RU }
\affiliation[122]{University of Jyvaskyla, FI-40014, Finland}
\affiliation[123]{Kansas State University, Manhattan, KS 66506, USA}
\affiliation[124]{Kavli Institute for the Physics and Mathematics of the Universe, Kashiwa, Chiba 277-8583, Japan}
\affiliation[125]{High Energy Accelerator Research Organization (KEK), Ibaraki, 305-0801, Japan}
\affiliation[126]{Korea Institute of Science and Technology Information, Daejeon, 34141, South Korea}
\affiliation[127]{K L University, Vaddeswaram, Andhra Pradesh 522502, India}
\affiliation[128]{National Institute of Technology, Kure College, Hiroshima, 737-8506, Japan}
\affiliation[129]{Taras Shevchenko National University of Kyiv, 01601 Kyiv, Ukraine}
\affiliation[130]{Lancaster University, Lancaster LA1 4YB, United Kingdom}
\affiliation[131]{Lawrence Berkeley National Laboratory, Berkeley, CA 94720, USA}
\affiliation[132]{Laborat{\'o}rio de Instrumenta{\c{c}}{\~a}o e F{\'\i}sica Experimental de Part{\'\i}culas, 1649-003 Lisboa and 3004-516 Coimbra, Portugal}
\affiliation[133]{University of Liverpool, L69 7ZE, Liverpool, United Kingdom}
\affiliation[134]{Los Alamos National Laboratory, Los Alamos, NM 87545, USA}
\affiliation[135]{Louisiana State University, Baton Rouge, LA 70803, USA}
\affiliation[136]{University of Lucknow, Uttar Pradesh 226007, India}
\affiliation[137]{Madrid Autonoma University and IFT UAM/CSIC, 28049 Madrid, Spain}
\affiliation[138]{Johannes Gutenberg-Universit{\"a}t Mainz, 55122 Mainz, Germany}
\affiliation[139]{University of Manchester, Manchester M13 9PL, United Kingdom}
\affiliation[140]{Massachusetts Institute of Technology, Cambridge, MA 02139, USA}
\affiliation[141]{Max-Planck-Institut, Munich, 80805, Germany}
\affiliation[142]{University of Medell{\'\i}n, Medell{\'\i}n, 050026 Colombia }
\affiliation[143]{University of Michigan, Ann Arbor, MI 48109, USA}
\affiliation[144]{Michigan State University, East Lansing, MI 48824, USA}
\affiliation[145]{Universit{\`a} del Milano-Bicocca, 20126 Milano, Italy}
\affiliation[146]{Universit{\`a} degli Studi di Milano, I-20133 Milano, Italy}
\affiliation[147]{University of Minnesota Duluth, Duluth, MN 55812, USA}
\affiliation[148]{University of Minnesota Twin Cities, Minneapolis, MN 55455, USA}
\affiliation[149]{University of Mississippi, University, MS 38677 USA}
\affiliation[150]{Universit{\`a} degli Studi di Napoli Federico II , 80138 Napoli NA, Italy}
\affiliation[151]{University of New Mexico, Albuquerque, NM 87131, USA}
\affiliation[152]{H. Niewodnicza{\'n}ski Institute of Nuclear Physics, Polish Academy of Sciences, Cracow, Poland}
\affiliation[153]{Nikhef National Institute of Subatomic Physics, 1098 XG Amsterdam, Netherlands}
\affiliation[154]{National Institute of Science Education and Research (NISER), Odisha 752050, India}
\affiliation[155]{University of North Dakota, Grand Forks, ND 58202-8357, USA}
\affiliation[156]{Northern Illinois University, DeKalb, IL 60115, USA}
\affiliation[157]{Northwestern University, Evanston, Il 60208, USA}
\affiliation[158]{University of Notre Dame, Notre Dame, IN 46556, USA}
\affiliation[159]{University of Novi Sad, 21102 Novi Sad, Serbia}
\affiliation[160]{Occidental College, Los Angeles, CA  90041}
\affiliation[161]{Ohio State University, Columbus, OH 43210, USA}
\affiliation[162]{Oregon State University, Corvallis, OR 97331, USA}
\affiliation[163]{University of Oxford, Oxford, OX1 3RH, United Kingdom}
\affiliation[164]{Pacific Northwest National Laboratory, Richland, WA 99352, USA}
\affiliation[165]{Universt{\`a} degli Studi di Padova, I-35131 Padova, Italy}
\affiliation[166]{Panjab University, Chandigarh, 160014 U.T., India}
\affiliation[167]{Universit{\'e} Paris-Saclay, CNRS/IN2P3, IJCLab, 91405 Orsay, France}
\affiliation[168]{Universit{\'e} Paris Cit{\'e}, CNRS, Astroparticule et Cosmologie, Paris, France}
\affiliation[169]{University of Parma,  43121 Parma PR, Italy}
\affiliation[170]{Universit{\`a} degli Studi di Pavia, 27100 Pavia PV, Italy}
\affiliation[171]{University of Pennsylvania, Philadelphia, PA 19104, USA}
\affiliation[172]{Pennsylvania State University, University Park, PA 16802, USA}
\affiliation[173]{Physical Research Laboratory, Ahmedabad 380 009, India}
\affiliation[174]{Universit{\`a} di Pisa, I-56127 Pisa, Italy}
\affiliation[175]{University of Pittsburgh, Pittsburgh, PA 15260, USA}
\affiliation[176]{Pontificia Universidad Cat{\'o}lica del Per{\'u}, Lima, Per{\'u}}
\affiliation[177]{University of Puerto Rico, Mayaguez 00681, Puerto Rico, USA}
\affiliation[178]{Punjab Agricultural University, Ludhiana 141004, India}
\affiliation[179]{Queen Mary University of London, London E1 4NS, United Kingdom }
\affiliation[180]{Radboud University, NL-6525 AJ Nijmegen, Netherlands}
\affiliation[181]{University of Rochester, Rochester, NY 14627, USA}
\affiliation[182]{Royal Holloway College London, TW20 0EX, United Kingdom}
\affiliation[183]{Rutgers University, Piscataway, NJ, 08854, USA}
\affiliation[184]{STFC Rutherford Appleton Laboratory, Didcot OX11 0QX, United Kingdom}
\affiliation[185]{Universit{\`a} del Salento, 73100 Lecce, Italy}
\affiliation[186]{San Jose State University, San Jos{\'e}, CA 95192-0106, USA}
\affiliation[187]{Sapienza University of Rome, 00185 Roma RM, Italy}
\affiliation[188]{Universidad Sergio Arboleda, 11022 Bogot{\'a}, Colombia}
\affiliation[189]{University of Sheffield, Sheffield S3 7RH, United Kingdom}
\affiliation[190]{SLAC National Accelerator Laboratory, Menlo Park, CA 94025, USA}
\affiliation[191]{University of South Carolina, Columbia, SC 29208, USA}
\affiliation[192]{South Dakota School of Mines and Technology, Rapid City, SD 57701, USA}
\affiliation[193]{South Dakota State University, Brookings, SD 57007, USA}
\affiliation[194]{Southern Methodist University, Dallas, TX 75275, USA}
\affiliation[195]{Stony Brook University, SUNY, Stony Brook, NY 11794, USA}
\affiliation[196]{Sun Yat-Sen University, Guangzhou, 510275}
\affiliation[197]{Sanford Underground Research Facility, Lead, SD, 57754, USA}
\affiliation[198]{University of Sussex, Brighton, BN1 9RH, United Kingdom}
\affiliation[199]{Syracuse University, Syracuse, NY 13244, USA}
\affiliation[200]{Universidade Tecnol{\'o}gica Federal do Paran{\'a}, Curitiba, Brazil}
\affiliation[201]{Tel Aviv University, Tel Aviv-Yafo, Israel}
\affiliation[202]{Texas A{\&}M University, College Station, Texas 77840}
\affiliation[203]{Texas A{\&}M University - Corpus Christi, Corpus Christi, TX 78412, USA}
\affiliation[204]{University of Texas at Arlington, Arlington, TX 76019, USA}
\affiliation[205]{University of Texas at Austin, Austin, TX 78712, USA}
\affiliation[206]{University of Toronto, Toronto, Ontario M5S 1A1, Canada}
\affiliation[207]{Tufts University, Medford, MA 02155, USA}
\affiliation[208]{Universidade Federal de S{\~a}o Paulo, 09913-030, S{\~a}o Paulo, Brazil}
\affiliation[209]{Ulsan National Institute of Science and Technology, Ulsan 689-798, South Korea}
\affiliation[210]{University College London, London, WC1E 6BT, United Kingdom}
\affiliation[211]{Valley City State University, Valley City, ND 58072, USA}
\affiliation[212]{Variable Energy Cyclotron Centre, 700 064 West Bengal, India}
\affiliation[213]{Virginia Tech, Blacksburg, VA 24060, USA}
\affiliation[214]{University of Warsaw, 02-093 Warsaw, Poland}
\affiliation[215]{University of Warwick, Coventry CV4 7AL, United Kingdom}
\affiliation[216]{Wellesley College, Wellesley, MA 02481, USA}
\affiliation[217]{Wichita State University, Wichita, KS 67260, USA}
\affiliation[218]{William and Mary, Williamsburg, VA 23187, USA}
\affiliation[219]{University of Wisconsin Madison, Madison, WI 53706, USA}
\affiliation[220]{Yale University, New Haven, CT 06520, USA}
\affiliation[221]{Yerevan Institute for Theoretical Physics and Modeling, Yerevan 0036, Armenia}
\affiliation[222]{York University, Toronto M3J 1P3, Canada}
\author[34]{A.~Abed Abud,}
\author[163]{B.~Abi,}
\author[67]{R.~Acciarri,}
\author[10]{M.~A.~Acero,}
\author[200]{M.~R.~Adames,}
\author[73]{G.~Adamov,}
\author[67]{M.~Adamowski,}
\author[19]{D.~Adams,}
\author[18]{M.~Adinolfi,}
\author[29]{C.~Adriano,}
% ...

\emailAdd{roberto@lbl.gov}

\abstract{The rapid development of general-purpose computing on graphics processing units (GPGPU) is allowing the implementation of highly-parallelized Monte Carlo simulation chains for particle physics experiments.
This technique is particularly suitable for the simulation of a pixelated charge readout for time projection chambers, given the large number of channels that this technology employs.
Here we present the first implementation of a full microphysical simulator of a liquid argon time {{projection}} chamber (LArTPC) equipped with light readout and pixelated charge readout, developed for the DUNE Near Detector. The software is implemented with an end-to-end set of GPU-optimized algorithms. The algorithms have been written in Python and translated into CUDA kernels using Numba, a just-in-time compiler for a subset of Python and NumPy instructions. The GPU implementation achieves a speed up of four orders of magnitude compared with the equivalent CPU version. The simulation of the current induced on $10^3$ pixels takes around $\SI{1}{\milli\second}$ on the GPU, compared with approximately $\SI{10}{\second}$ on the CPU.
The results of the simulation are compared against data from a pixel-readout LArTPC prototype.
}
\arxivnumber{2212.09807}
\begin{document}
\maketitle
\flushbottom

\section{Introduction}
\label{sec:intro}

The idea of using a liquid argon time projection chambers (LArTPC) for the detection of neutrino interactions was first proposed in 1977 \cite{Rubbia:1977zz}. The detection mechanism is the following: charged particles produced by neutrino interactions ionize the argon, leaving a trail of ionization electrons. In addition, liquid argon also produces scintillation light, which provides calorimetric information and a fast timing signal ($\mathcal{O}(10~\mathrm{ns})$ \cite{hitachi:1983}). A fraction of the ionized electrons recombine immediately with the positive argon ions, while the remaining ones drift towards the anode side of the detector in a homogeneous electric field applied to the argon volume, which is usually $\mathcal{O}(100~\mathrm{V/cm})$. Impurities present in the LAr (e.g. O$_\mathrm{2}$, H$_\mathrm{2}$O, N$_\mathrm{2}$) can attach a portion of the drifting electrons. The amount of drifting electrons declines as a function of the distance from the anode, since the electrons need to travel a longer path. 

Typically, two or more arrays of sense wires are placed at the anode and assembled into planes. The drifting of negative charges in a constant electric field induces a signal on the wires. Each plane provides a two-dimensional image of the ionization: the position of the wire provides one dimension, and the time of the arrival provides the second one, since the drift velocity of the electrons in the LAr is known and is typically $\mathcal{O}(1~\mathrm{m/ms})$. Using multiple wire planes can help estimate the position of the ionization in three dimensions. However, ambiguities arise when drifting electrons are isochronous or parallel to a wire orientation. Unambiguous 3D imaging of LArTPC charge signals is possible using a readout system based on a pixelated array of charged-sensitive pads, which has been demonstrated in ref. \cite{Dwyer:2018phu}. Both the typical distance between adjacent wires and the typical pixel pitch are in the order of few millimeters. The position on the anode plane of the involved pads provides two spatial dimensions, and the time of the induced signal provides the third one. This truly three-dimensional readout provides better reconstruction efficiency and purity than the 2D combined wire readout, as demonstrated in ref.~\cite{Adams:2019uqx}.

Pixel readout requires the channel count used to be increased by a factor of 10 to 100 with respect to wire planes. Thus, with this increased channel count and granularity simulation burden, the transport of electrons in LAr and the signal induction on the pixel pads represent an ideal use case for highly-parallelized, concurrent simulation algorithms. The development of general-purpose computing on graphics processing units (GPGPU), which went hand in hand with the advances in the machine learning and deep learning fields, has driven the design of the current generation of supercomputers, such as Perlmutter at NERSC \cite{yang2020accelerate}, which is a heterogeneous system with both GPU-accelerated and CPU-only nodes. Implementing highly-parallelized simulation chains allows full advantage of these new systems to be taken and enables the simulation of future pixelated LArTPCs, which would otherwise not be viable with current resources.

The US-based neutrino physics program relies on present and future experiments using LArTPC technology. The flagship experiment is the Deep Underground Neutrino Experiment (DUNE), which will consist of a high-intensity accelerator neutrino beam, measured by near and far detectors \cite{Abi:2020wmh}. The Far Detector will consist of four 17-kiloton LArTPC modules located deep underground at the Sanford Underground Research Facility in Lead, South Dakota, located 1285 km from the beam source \cite{Abi:2020evt}. The Near Detector will be located at Fermilab, 574 m from the beam source, and will contain a 67 t modular LArTPC called ND-LAr \cite{DUNE:2021tad}.

There are already some software toolkits for LArTPCs \cite{Snider:2017wjd, Szydagis:2011tk} and there has been some effort towards parallelizing the reconstruction stage \cite{Berkman:2021ffy}. However, the simulation stages have remained mostly sequential and their speed up has been recognized as a priority by the community \cite{Boyle:2022cvo}.

In this document we will describe the implementation of a set of highly-parallelized algorithms, organized in a module called \texttt{larnd-sim} \cite{larndsim}, that run on GPUs. They simulate the ionized electrons recombination and drifting towards the anode, the generation of electronics signals on the pixelated readout, and the processing of the signal by the front-end electronics.

\section{Technical implementation}
Recent rapid developments in the field of machine learning have stimulated the creation of several tool-kits for GPU-accelerated applications. In particular, the NVIDIA\textregistered~CUDA platform \cite{CUDA} allows to use GPUs for general purpose computing via different programming languages. We opted for Numba \cite{numba}, which generates CUDA computing kernels using a subset of native Python and NumPy code \cite{numpy}.

A CUDA kernel is a function that is executed $N$ times in parallel by $N$ CUDA threads. The threads can be organized in one-dimensional, two-dimensional, or three-dimensional \emph{blocks}, which in turn can be organized in one-dimensional, two-dimensional, or three-dimensional \emph{grids}. Blocks in the same grid contain the same number of threads, can run independently, and can be executed in any order, while threads in the same block can co-operate through shared memory.
CUDA kernels typically store the result of the computation in a pre-allocated array passed to the kernel function. The CUDA programming model requires a careful design of the algorithm: the shape and size of the array where the result is stored must be known in advance and the threads must avoid race conditions during execution\footnote{In software, a race condition can happen when the behavior of the program depends on the relative timing of multiple threads or processes.}, thus the result of the algorithm must not depend on the order of execution of the threads.

The software described in this document contains several CUDA kernel functions, separated into two logical categories: one for the charge simulation, described in section \ref{sec:charge}, and one for the light simulation, described in section \ref{sec:light}. The functions simulate the detector response, including: (1) the recombination of the electrons with the argon ions, (2) the drifting of the electrons towards the anode, (3) the induction of electronic signals on the pixel pads and optical detectors, and (4) the electronics response of the charge and light readout systems.

\begin{figure}[htbp]
\centering
\includegraphics[width=0.98\textwidth]{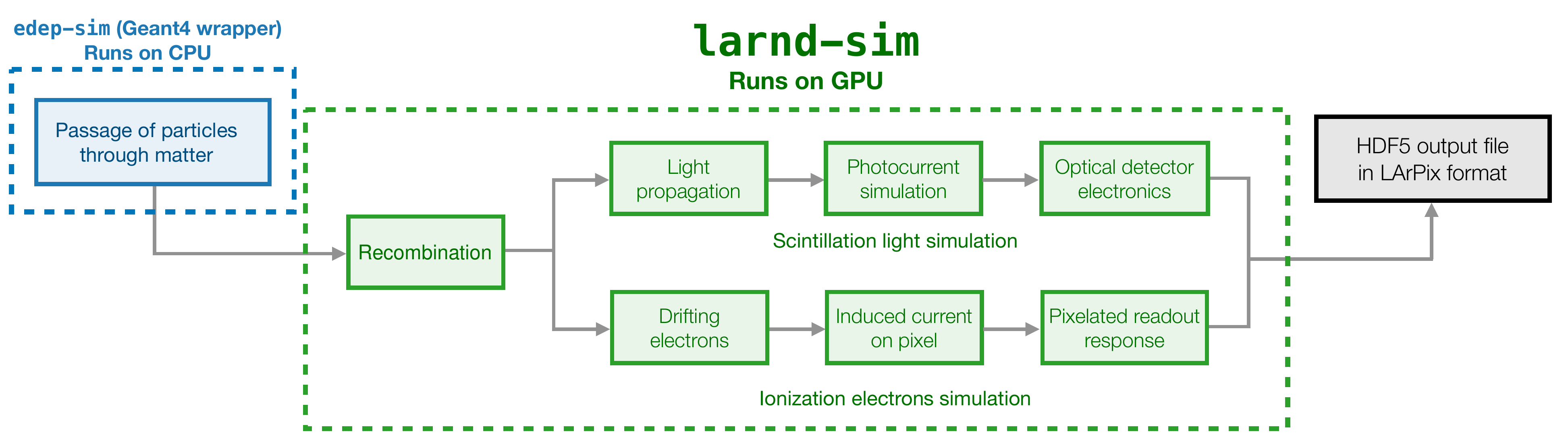}
\caption{Diagram showing the full simulation workflow. The passage of the particle through matter is simulated by \texttt{edep-sim} on the CPU. The output is fed to \texttt{larnd-sim}, which runs entirely on GPU. Its output is finally saved in a HDF5 file.} \label{fig:larndsim_diagram}
\end{figure}

The simulation of the passage of the initial particles through matter is performed using \texttt{edep-sim} \cite{edepsim}, a wrapper around \textsc{Geant4} \cite{Agostinelli:2002hh}, which is independent from the \texttt{larnd-sim} package described here. The output consists of a set of short particle track segments, in the order of few millimeters, which describe the energy deposition trail of each particle. The length of the segments depends on the derivative of the stopping power $dE/dx$: the portion of a particle trail where the $dE/dx$ changes abruptly will be divided in finer segments than the portion where the $dE/dx$ is mostly constant.  Thus, in a single segment, the energy deposition per unit length is assumed to be constant. {{This approximation is valid when the segment length is of the same order as the detector resolution -- at much shorter lengths, fluctuations from the long tail of the dE/dx distribution fall outside of the applicable region of recombination models, and at much longer lengths, correlations in the dE/dx fluctuations become significant and the dE/dx width will be under-simulated. Within these broad considerations, reducing the minimum size of the segments has not shown a significant impact on the result of the charge simulation}}. This set $S_{i}$ is stored in a bi-dimensional NumPy array containing the energy deposition and the spatial distribution of the segments:
\begin{equation}
    S_{i} = \left(\vec{r_s}, \vec{r_e}, E\right)_i,
\end{equation}
where $\vec{r_s}$ and $\vec{r_e}$ are four-dimensional vectors containing the spatial and timing coordinates of the segment start and end points, and $E$ is the deposited energy.
This array is used as input for our module. In order to minimize the memory transfer between the host and the device (in our case the GPU), we allocate the NumPy array directly on the device memory using CuPy \cite{cupy}, a GPU array backend that implements a subset of the NumPy interface. The output of the \texttt{larnd-sim} simulation is then saved in a HDF5 file \cite{hdf5}. The entire simulation workflow is shown in figure \ref{fig:larndsim_diagram}.

\section{Charge simulation}\label{sec:charge}

\subsection{Electron recombination}\label{sec:reco}
The first step of our simulation is to calculate the number of electrons that remain after the recombination and start drifting towards the anode. We denote the initial charge ionized by the particle as $Q_0$, while the charge remaining after the recombination is given by $Q_R = \mathscr{R}\cdot Q_0$, where $\mathscr{R}$ is our recombination factor.

Two different models are commonly used to describe this phenomenon: the Birks model \cite{Birks:1951boa}, which gives spurious values when applied to high-ionization particles \cite{Acciarri:2013met}, and the modified Box model \cite{Acciarri:2013met}, which doesn't suffer from these issues but is inadequate to describe particles at low stopping power (low $dE/dx$).

The recombination factor for the Birks model $\mathscr{R}_{\mathrm{Birks}}$ can be parametrized as:
\begin{equation}\label{eq:birks}
    \mathscr{R}_{\mathrm{Birks}} = \frac{A_b}{1+k_b/\epsilon\cdot dE/dx },
\end{equation}
where $A_b$ and $k_b$ are free parameters that usually depend on the detector and $\epsilon$ is the product of the electric field with the liquid argon density. The ICARUS collaboration obtained $A_b=0.800$ and $k_b=0.0486$~(kV/MeV)~(g/cm$^3$) \cite{ICARUS:2004koz}, which are the values used in our simulation. 

The recombination factor the modified Box model $\mathscr{R}_{\mathrm{Box}}$ is defined as:
\begin{equation}\label{eq:box}
    \mathscr{R}_{\mathrm{Box}} = \frac{\log(\alpha + \beta/\epsilon\cdot dE/dx)}{\beta/\epsilon\cdot dE/dx},
\end{equation}
where $\alpha$ and $\beta$ are free parameters which were measured by the ArgoNeuT collaboration to be $\alpha=0.93$ and $\beta=0.207$~(kV/MeV)(g/cm$^3$) \cite{Acciarri:2013met}.
In both cases, the typical recombination factor for a minimum ionizing particle (MIP) is around 0.7. Our simulation assumes the Birks model by default. 



The implementation of the calculation of the recombination factors $\mathscr{R}_{\mathrm{Birks}}$ or $\mathscr{R}_{\mathrm{Box}}$ on the GPU is trivial: the $i$-th thread of the $K_\mathrm{recomb}(S_i, \mathscr{R})$ CUDA kernel takes as input the $i$-th row of the NumPy array containing the segments $S$ and applies the recombination formula, so eq. \eqref{eq:birks} or eq. \eqref{eq:box}. The result is stored in an appropriate column of the NumPy array. {Currently, the simulation treats the recombination factor as a fixed number: although this is an approximation, the level of fluctuations associated with it is significantly lower than the expected intrinsic noise of the detector, which was measured with the prototype described in section \ref{sec:data}}.

\begin{figure}[htbp]
\centering
\includegraphics[width=0.7\textwidth]{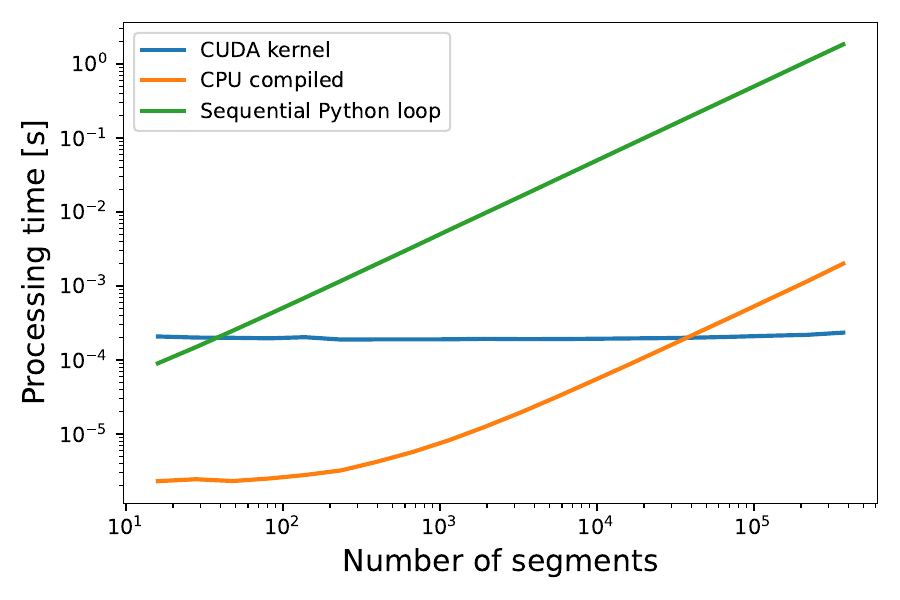}
\caption{Processing time for the calculation of the recombination factor $\mathscr{R}_{\mathrm{Birks}}$ of eq. \eqref{eq:birks} with the GPU implementation using a CUDA kernel (blue), the CPU implementation using Numba (orange), and a sequential Python \texttt{for} loop (green). For reference, a neutrino beam spill corresponds on average to $\mathcal{O}(10^5)$ energy deposition segments in the ND-LAr. This computation was performed on a node of the NERSC Cori supercomputer, which contains two sockets of 20-core Intel Xeon Gold 6148 (Skylake) at 2.40 GHz and 8 NVIDIA\textregistered~Tesla\textregistered~V100 (Volta) GPUs.}\label{fig:comparison}
\end{figure}

Even if this operation is computationally inexpensive, it's instructive to take a look at the performance comparison between a sequential, interpreted Python \texttt{for} loop, a loop compiled on the CPU using Numba, and the GPU implementation using a CUDA kernel. Figure \ref{fig:comparison} shows the processing time needed to calculate $\mathscr{R}_{\mathrm{Birks}}$ of eq. \eqref{eq:birks} as a function of the number of \textsc{Geant4} segments given as input.
The CPU-compiled version is obviously faster than the sequential interpreted loop since the function is now translated into machine code. While the CUDA kernel processing time is initially the largest, it doesn't immediately scale with the number of input segments, so it starts being the exponentially faster implementation with more than $\mathcal{O}(10^4)$ segments, where it starts taking advantage of the massive parallelization achievable by the GPU. The NVIDIA\textregistered~Tesla\textregistered~V100 GPU used for this study can run more than $10^5$ parallel threads. To give a figure of merit, a typical neutrino beam spill in the ND-LAr corresponds on average to $\mathcal{O}(10^5)$ segments.

\subsection{Electron transport in liquid argon}\label{sec:drifting}
The electrons remaining after the recombination travel towards the anode at a constant velocity $v_\mathrm{drift}$ (assuming perfect uniformity of the electric field), which is typically $\mathcal{O}(1~\si{\milli\meter}/\si{\micro\second})$. The time they take to reach the anode is given by:
\begin{equation}
    t_{\mathrm{drift}} = (z - z_{\mathrm{anode}})/v_\mathrm{drift},\label{eq:drift}
\end{equation}
where the electron drift direction is assumed to be along the $z$ axis and where $z_\mathrm{anode}$ is the $z$ coordinate of the anode.

The impurities in the liquid argon, such as O$_2$, N$_2$ and H$_2$O, can attach a portion of the drifting electrons, so electrons farther from the anode will have a higher chance to be attached. This effect is usually parametrized by a negative exponential, so the charge $Q_a$ that effectively reaches the anode, assuming uniform impurities and a perfect electronics response, is:
\begin{equation}
    Q_a = Q_R\cdot\exp(-t_{\mathrm{drift}}/\tau),\label{eq:lifetime}
\end{equation}
where $t_{\mathrm{drift}}$ is the time the charge takes to reach the anode and $\tau$ is a parameter that depends on the concentration of impurities and it is usually called \emph{electron lifetime}, which is in the order of milliseconds for concentrations of O$_2$ at tens of parts per trillion.

Electrons drifting in strong electric fields do not diffuse isotropically, so it is necessary to estimate both the longitudinal and transverse components with respect to the drift direction \cite{Li:2015rqa}. The diffusion length is given by:
\begin{equation}
    \sigma = \sqrt{2D t_{\mathrm{drift}}},
\end{equation}
where $D$ is the longitudinal or transverse diffusion coefficient, which depends on the electric field and liquid argon temperature. In our simulation we set the longitudinal and transverse diffusion coefficients to $D_l=4$~cm$^2$/s and $D_t=8.8$~cm$^2$/s, respectively. These values were obtained by a preliminary ProtoDUNE-SP \cite{DUNE:2021hwx} analysis. 

This information is calculated and stored in appropriate columns of the NumPy array in a way analogous to the one described in section \ref{sec:reco}, where each thread processes a single segment independently.

\subsection{Electronic signal induction on a pixel}
\subsubsection{Field response}\label{sec:field}

The current induced by a point charge on a given pixel within the anode is calculated using the Shockley-Ramo theorem \cite{Ramo:1939vr}:
\begin{equation}
    I_{\mathrm{pixel}}
    = q \; \vec{v} \cdot \nabla W,\label{eq:ramo}
\end{equation}
where $W$ is the \emph{weighting field {potential}}, the normalized contribution of a single electrode to the overall field, $q$ is the particle charge, and $\vec{v}$ is the velocity of the particle.  
Within \texttt{larnd-sim}, the induced current is pre-calculated for a point-like charge and referenced as a $I_{\mathrm{pixel}}[t, x, y]$ look-up table (LUT). The table contains the current induced by an electron placed at discrete $(x,y)$ locations on the anode plane at a discrete time $t$, where $t=0$ is the time when the electron is at $z=0.5$~cm. This value is chosen because of the observed flatness of the electric and weighting fields at this distance. 

\begin{figure}[htbp]
     \centering
     \begin{subfigure}[b]{0.45\textwidth}
         \centering
         \includegraphics[width=\textwidth]{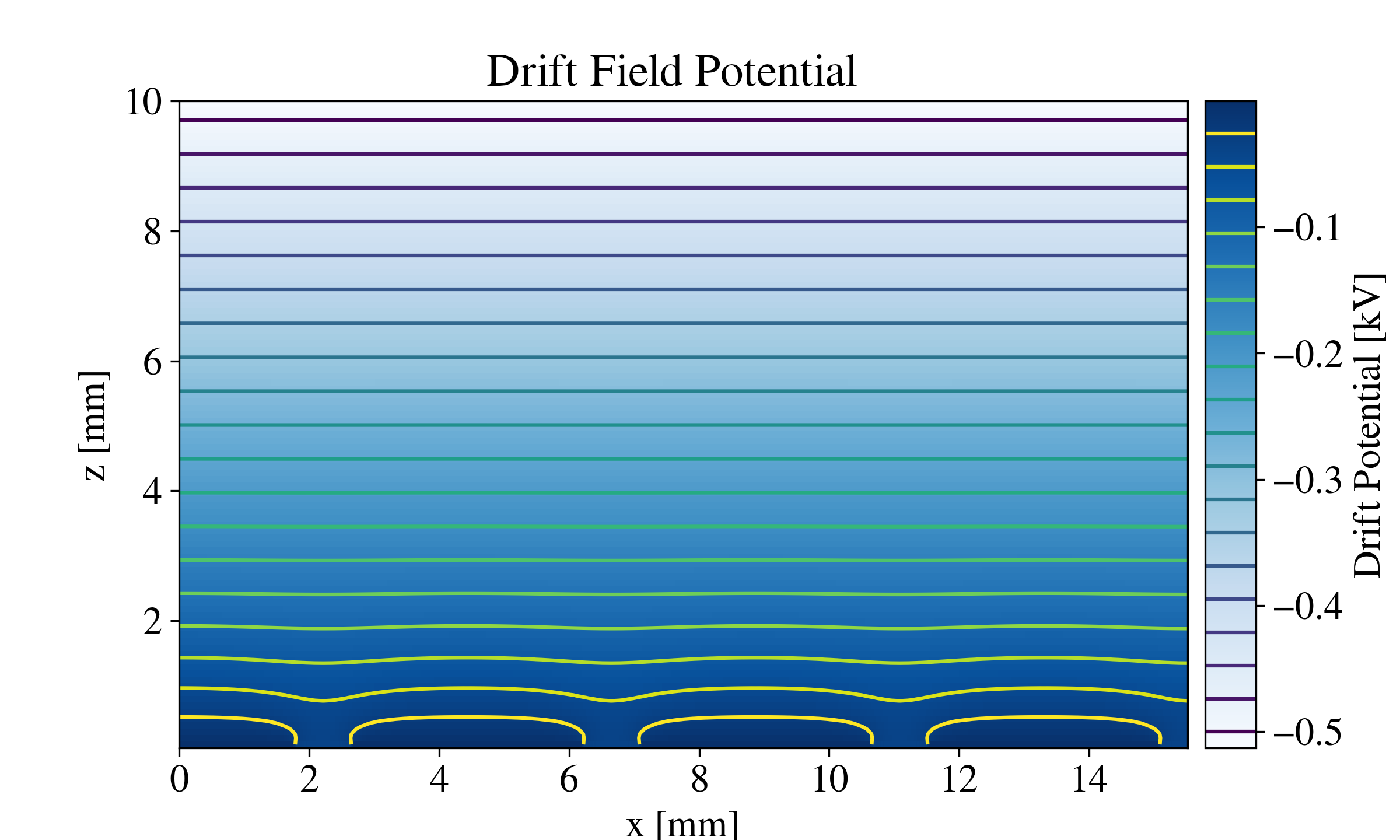}
         \caption{Drift electric potential field in the region near the pixel plane.  Shown is a slice in $y$ along on a pixel center.  The pixel surfaces are set to 0 kV, while the gradient of the far field at the surface of $z = 10$~mm is set to 0.5~kV/cm. The color scale corresponds to the value of the drift potential.}
     \end{subfigure}
     \hspace{2.5em}
     \begin{subfigure}[b]{0.45\textwidth}
     \raisebox{20mm}
         \centering
         \includegraphics[width=\textwidth]{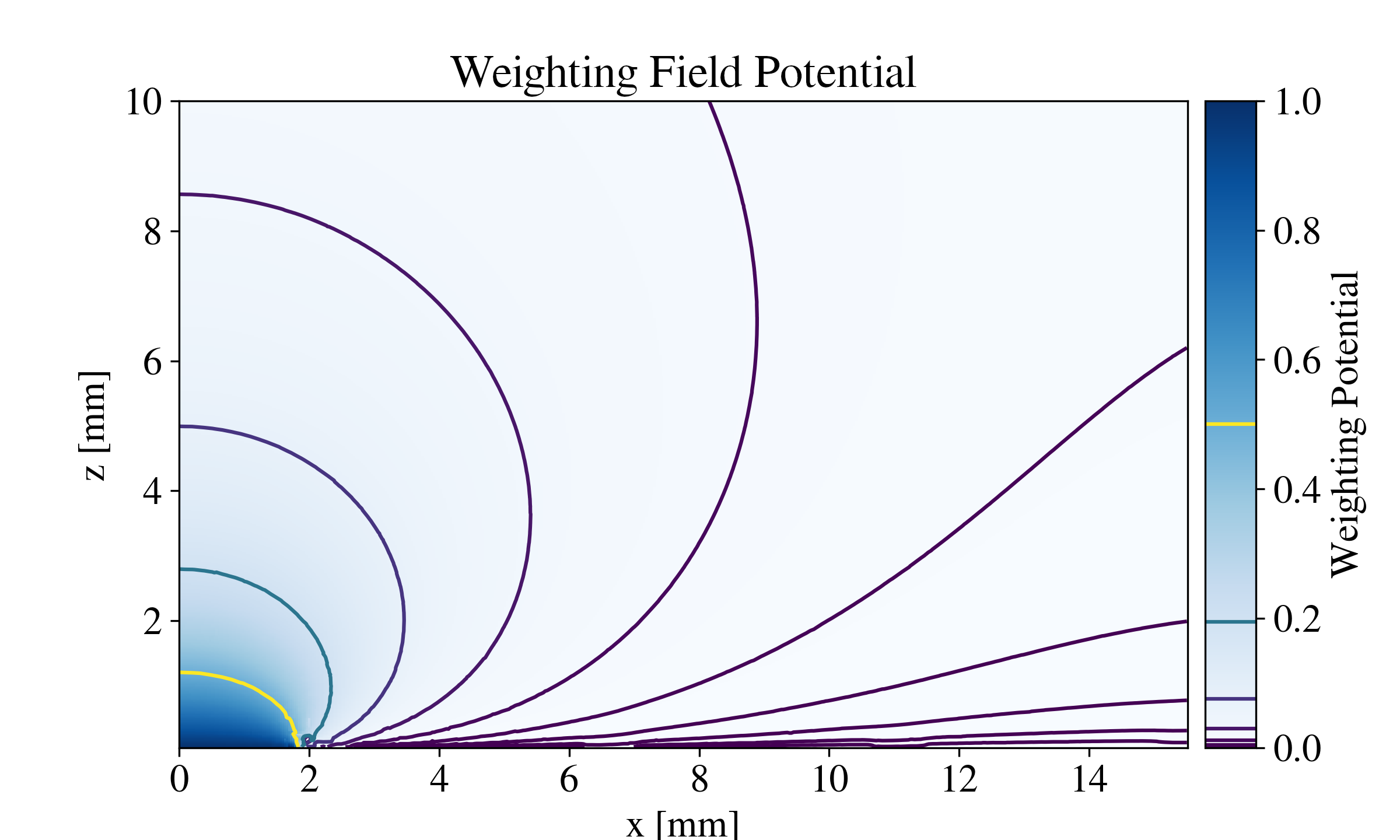}
         \caption{Weighting potential field in the region near a "pixel of interest".  The central ($x = y = 0$ pixel is set to a unit potential, while all other surfaces are set to 0 potential.  The resulting unitless field defines the susceptibility to current induction by charges moving nearby.}
     \end{subfigure}
     \caption{The drift (left) and weighting (right) {{field potentials}} obtained by finite-element analysis in the region near a generic pixel.  Shown are slices of the potential fields along an $x-z$ plane which crosses through the centerline of a pixel.}
     \label{fig:fieldMaps}
\end{figure}

The $I_{\mathrm{pixel}}[t,x,y]$ values are calculated as follows. In the region very close to the anode, the electric field and the weighting field are calculated numerically. The geometry of a small volume, including a central pixel with a pitch of 4.4~mm and its 8 nearest neighbor pixels, is modeled using CAD software and converted into a 3-dimensional mesh using the Gmsh package \cite{Gmsh}. The fields are then calculated using the successive over-relaxation method as implemented in the Elmer FEM \cite{elmer} software package. The electric and weighting fields share the same geometry, but differ in the boundary conditions imposed on the problem. In the electric field calculation, the pixels are grounded, with the backing plane at a small offset voltage, and the field on the cathode-facing side of the volume is set to the nominal field of 500~V/cm. In the calculation of the weighting field, all electrodes other than the central pixel are grounded, the central pixel is set to unit voltage, and the voltage on the remaining conductors is set to zero.  The results of these two field calculations are shown in figure \ref{fig:fieldMaps}.

\begin{figure}[htbp]
     \centering
     \begin{subfigure}[b]{0.45\textwidth}
         \centering
         \hspace{-4em}
         \includegraphics[width=1.1\textwidth]{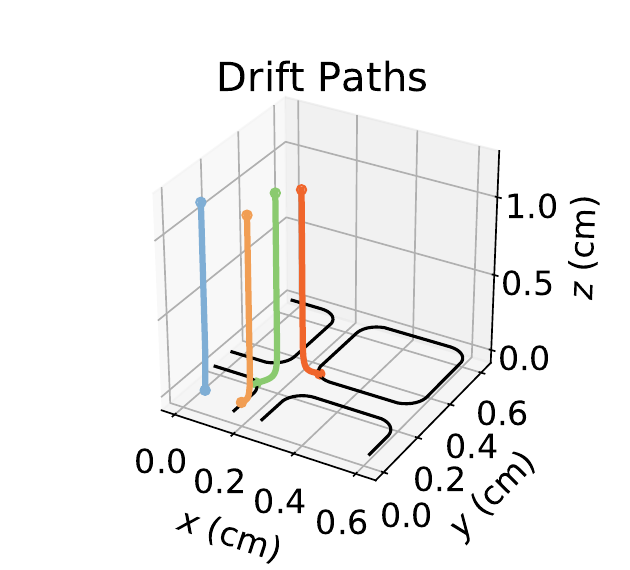}
         \caption{Example drift paths for electrons starting at a short distance from a pixel centered at $(0,0).$}
         \label{fig:driftpaths}
     \end{subfigure}
     \hspace{2.5em}
     \begin{subfigure}[b]{0.45\textwidth}
     \raisebox{20mm}
         \centering
         \includegraphics[width=\textwidth]{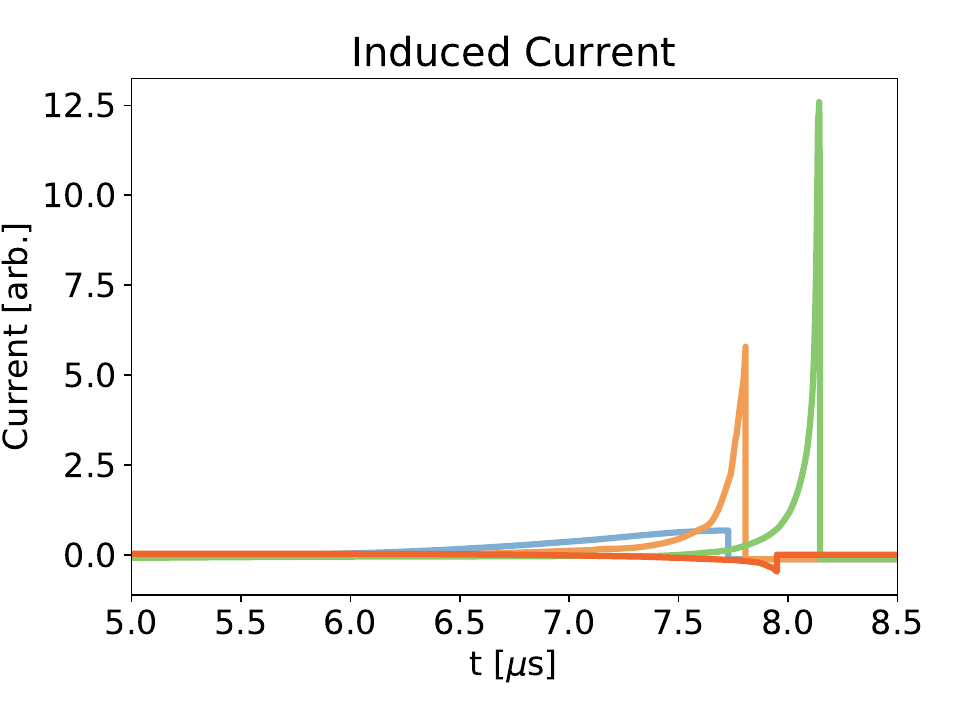}
         \caption{Induced current on the pixel at $(0,0)$ from example drift paths}
         \label{fig:driftpaths_current}
         \vspace{1em}
     \end{subfigure}
     \caption{To determine the near-field pixel response, we simulate the paths and velocities of drift electron within a 3D FEM field model. Here we show the simulated electron drift paths (left), and the corresponding induced current on the pixel (right) for a subset of the starting points used. The color of the trajectory on the left corresponds to the color of the current induced curve on the right.}
\end{figure}

Next, the idealized drift paths (ignoring diffusion and attenuation effects) are integrated using the ICARUS and Walkowiak \cite{Amoruso:2004ti, Walkowiak:2000wf} electron transport models evaluated within the calculated electric field.  This model allows for the drift velocity to change as a function of the local electric field, as the assumption of a perfectly uniform field is not necessarily true very close to the anode. The drift paths are calculated for a grid of four hundred $(x,y)$ positions, 0.5~cm from the anode on the $z$ axis. The resulting granularity on the $x$ and $y$ axes is 0.33~mm. Finally, we compute the time-derivative of the weighting field {{potential}} along each path, which yields a charge-normalized current series for a given initial charge position. The sampling in time is $0.1$~\si{\micro\second}. A select few drift paths are shown in figure~\ref{fig:driftpaths}, with their corresponding current series in figure~\ref{fig:driftpaths_current}.

\begin{figure}[htbp]
\centering
\includegraphics[width=0.9\textwidth]{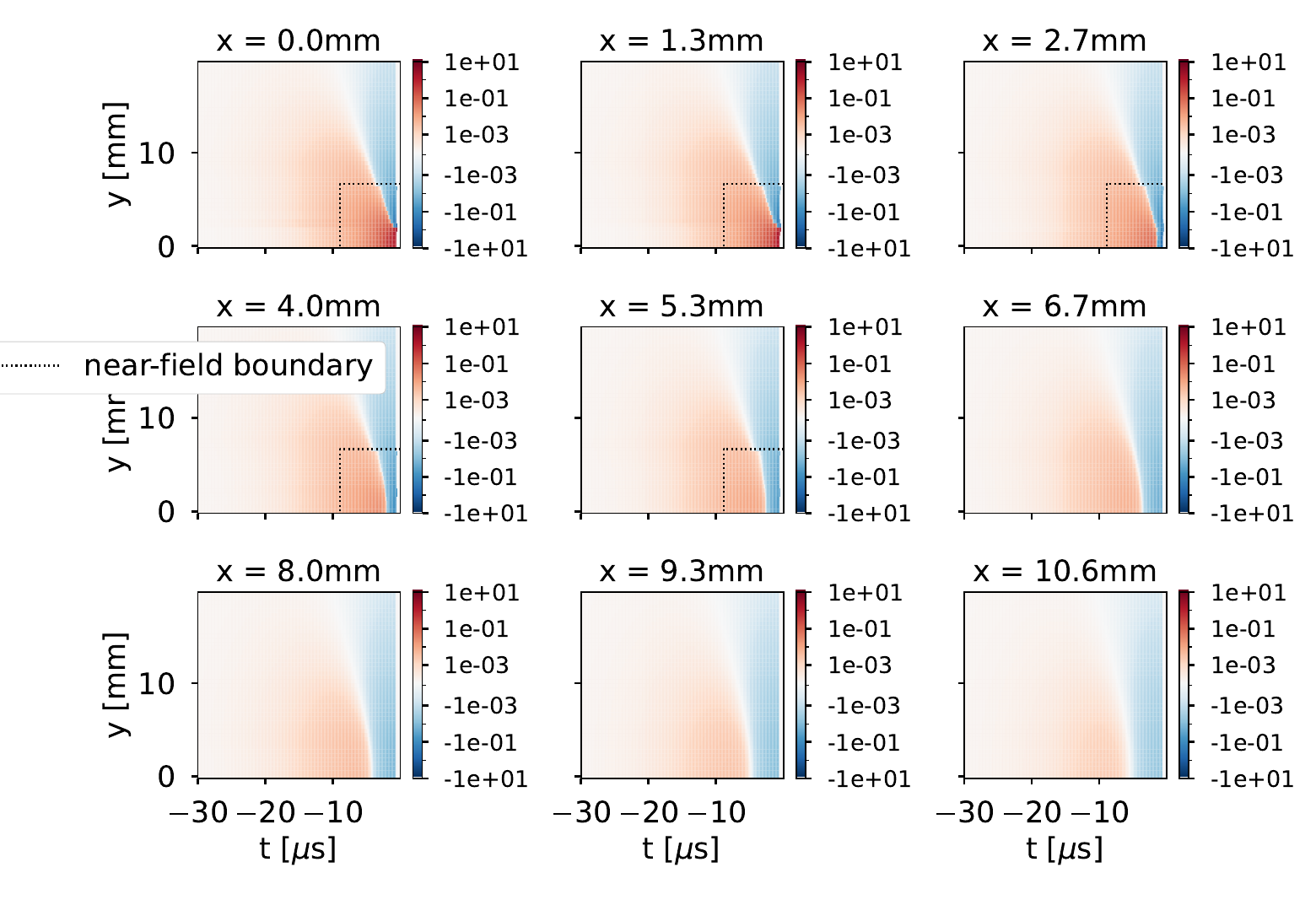}
\caption{Induced current on a 4.4mm pixel from a drifting electron at a given drift time and $x, y$ position relative to the pixel, in units of electrons per $0.1~$\si{\micro\second}. The different calculations used for the near-field and far-field regions are described in the text in section~\ref{sec:field} with the near-field boundary highlighted with a dotted line.} \label{fig:fieldresponse}
\end{figure}

Farther from the pixel at $z>0.5$~cm, the weighting potential is solved by treating the pixel as point-like and using a method-of-image-charges approximation to fix the boundary conditions at the anode and the cathode. The drift velocity is assumed to be uniform. The relative normalization to the near-field calculation is then fixed by enforcing continuity in the current series at the time boundary. Some re-scaling is performed on these current series to ensure they integrate to one electron charge, as error is introduced in the drift path integration, as well as numerical error present in the FEM solutions and their interpolated values. Figure \ref{fig:fieldresponse} shows the tabulated induced current on a pixel up to $\SI{30}{\micro\second}$ from the time of arrival of the drift electron. The agreement of the near-field FEM model and the dipole far-field approximation across this surface seen in figure \ref{fig:fieldresponse} demonstrates that the transition surface at $0.5$~cm is sufficiently far from the pixel plane.

\subsubsection{Induced current calculation}\label{sec:induced}
The electrons drifting towards the anode are diffused in the longitudinal and transverse directions, as described in section \ref{sec:drifting}. The ionization electrons corresponding to a \textsc{Geant4} segment will then form a three-dimensional charge cloud. The induced current on a given pixel from this can be calculated by taking the convolution of the pre-calculated pixel response model (see section~\ref{sec:field}) and the charge density of the track segment:
\begin{align}
    I(t) &= \int I_\mathrm{pixel}(t-z/v_d,x,y) \rho(x,y,z) dx dy dz
\end{align}
where $\rho(x,y,z)$ is the 3-dimensional charge density including diffusion and $I_\mathrm{pixel}(t-z/v_d,x,y)$ is the current response model at the given position and time tick, which is stored in the LUT described in section \ref{sec:field}. The positively charged ions also produce a current on the pixel, however the magnitude of this current is more than 7 orders of magnitude smaller than the signal from the drift electrons due to the small drift velocity of the ions, and so it is neglected in the detector simulation.

The calculation of $I(t)$ is the most computational-intensive step of the detector simulation. In order to reduce its complexity, we apply two approximations to the charge density: first, we discretize the track segment along the track length into $N$ points; and second, we approximate the effect of diffusion summing the contribution from $M$ random 3D perturbations of the sample point. With these approximations, we can write the charge density as
\begin{align}
    \rho(x,y,z) &\approx \frac{Q_a}{N M} \sum_{\mu=1}^M \sum_{\nu=1}^N \delta(x_\mu+\delta x_{\mu\nu} - x) \delta(y_\mu+\delta y_{\mu\nu} - y) \delta(z_\mu+\delta z_{\mu\nu} - z)
\end{align}
where $Q_a$ is the total charge magnitude of the segment from eq. \eqref{eq:lifetime}, $\delta(x)$ is the Dirac delta function, $N$ is the number of sampled points along the track segment, $M$ is the number of Monte Carlo iterations per sample point, $(x_\mu,y_\mu,z_\mu)$ is a point uniformly distributed along the track segment, and $(\delta x_{\mu\nu}, \delta y_{\mu\nu}, \delta z_{\mu\nu})$ are normally distributed in proportion to the diffusion in that dimension. This simplifies the calculation to a sum over the pixel response model at each sampled position
\begin{align}
    I(t) &\approx \frac{Q_a}{NM} \sum_{\mu=1}^N \sum_{\nu=1}^M I_\mathrm{pixel}\left(t-(z_\mu + \delta z_{\mu\nu})/v_d, x_\mu + \delta x_{\mu\nu}, y_\mu + \delta y_{\mu\nu}\right).\label{eq:current}
\end{align}
In our implementation, $N$ is calculated for each track with a $\SI{10}{\micro\meter}$ sampling interval, with the option of increasing the resolution and the Monte Carlo sampling for higher precision. For reference, at this sampling interval each sample point contributes a charge of roughly 50~electrons for a MIP-like track, and the error introduced by the approximations is at the percent level and much smaller than the noise level.

The calculation of the induced current is particularly well-suited to be implemented on a highly-parallelized architecture, since it is necessary to perform the sum of eq. \eqref{eq:current} for each time tick in the time window, each pixel, and each \textsc{Geant4} detector segment. The length of the time window depends on the orientation of the segment with respect to the anode, since the arrival time is directly proportional to the $z$ coordinate. In order to minimize the memory usage, the length of the longest signal and the largest number of pixels per segment are calculated before execution and used to allocate the array where the induced signals are going to be stored.

In our CUDA kernel $K_\mathrm{current}(s_i, p_j, t_k)$, the parallel threads are organized in a three-dimensional grid, where the dimensions are the segment index $s_i$, the pixel index $p_j$, and the time tick $t_k$.

\begin{figure}[htbp]
\centering
\includegraphics[width=0.7\textwidth]{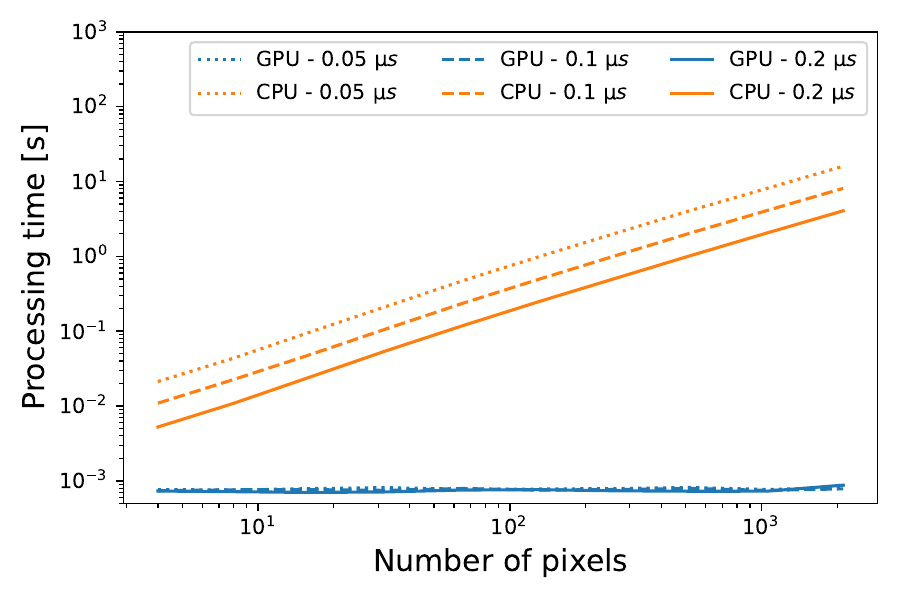}
\caption{Processing time for the calculation of the induced current with the GPU implementation using a CUDA kernel (blue) and with a CPU implementation compiled with Numba with explicit parallelization of the loops (orange). For reference, a neutrino beam spill will induce a signal on around $5\times10^4$ pixels of the ND-LAr. The calculation has been performed with different values of the time sampling ({{$\SI{0.05}{\micro\second}$, $\SI{0.1}{\micro\second}$, $\SI{0.2}{\micro\second}$}}) on a node of the NERSC Cori supercomputer, which contains two sockets of 20-core Intel Xeon Gold 6148 (Skylake) at 2.40 GHz and 8 NVIDIA\textregistered~Tesla\textregistered~V100 (Volta) GPUs.}\label{fig:cpugpu}
\end{figure}

It is possible to compare the performances of this GPU algorithm with an equivalent CPU version, compiled with explicit parallelization of the loops using the Numba \texttt{prange} function. This function runs the loop in parallel threads, one for each core (20 in our case), and tries to merge adjacent loops together, reducing loop overhead \cite{parallelcomputing}. The GPU version provides a speed-up of around four orders of magnitude for $\mathcal{O}(10^3)$ simulated pixels, as shown in figure \ref{fig:cpugpu} for different values of the time sampling.
Interestingly, in this case the GPU version is significantly faster than the CPU one even for only 10 pixels, since we are parallelizing also in two other dimensions (segment index and time).
The processing time for the GPU implementation does not increase with the number of simulated pixels and with different values of the time sampling, which shows that the calculation is effectively being performed in parallel. 
Compilation times and memory allocation times are not taken into account here. 

\begin{figure}[htbp]
     \centering
     \begin{subfigure}[b]{0.3\textwidth}
         \centering
         \includegraphics[width=\textwidth]{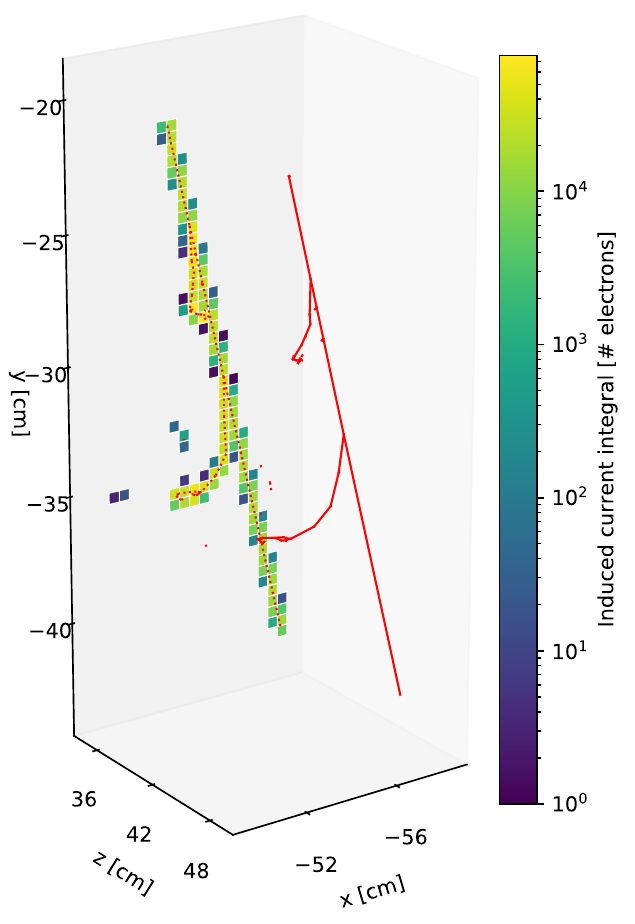}
         \caption{3D event display of a simulated cosmic-ray muon.}
         \label{fig:evdexample}
     \end{subfigure}
     \hspace{2.5em}
     \begin{subfigure}[b]{0.5\textwidth}
     \raisebox{20mm}
         \centering
         \includegraphics[width=\textwidth]{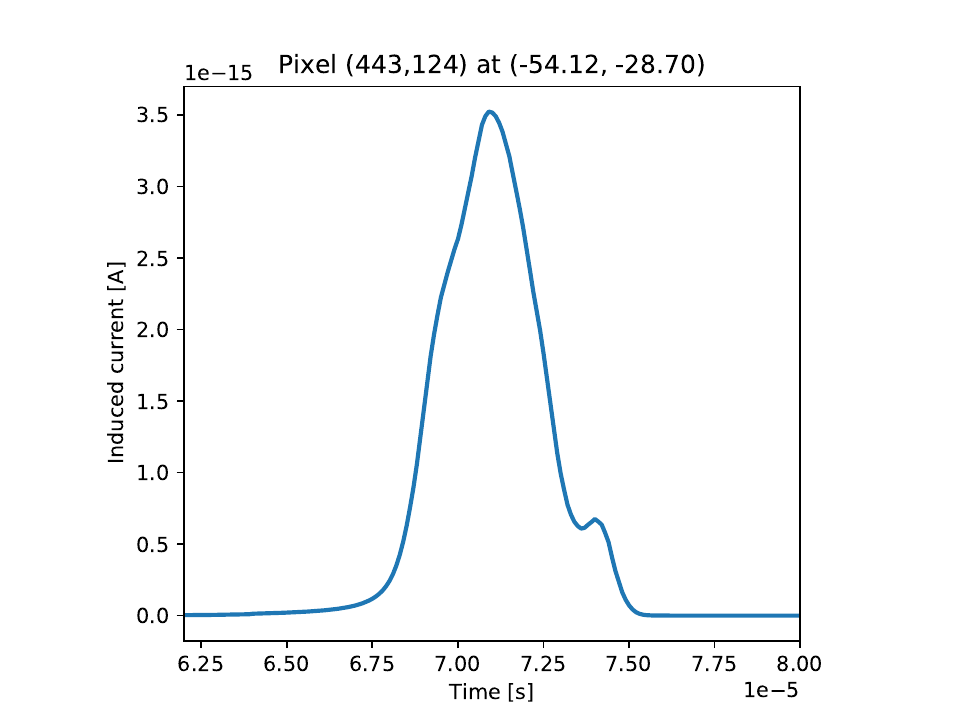}
         \caption{Current induced on a single pixel by the cosmic-ray muon and a delta ray.}
         \label{fig:evdcurrent}
     \end{subfigure}
     \caption{The simulation of the induced current performs the calculation of eq. \eqref{eq:current} for each active pixel. Here we show a 3D event display of a simulated cosmic-ray muon (left, in red) and the current induced on a pixel by the cosmic-ray muon and a delta ray (right).}
\end{figure}

The result of the simulation of a cosmic-ray muon interacting in the TPC active volume is shown in figure \ref{fig:evdexample}. Given the effect of transverse diffusion, signals are induced also on the pixels that do not lie exactly on the $xy$ projection of the muon trajectory. The high granularity of the pixel layout shows clear imaging of features such as delta rays. Figure \ref{fig:evdcurrent} shows the current induced on a pixel both by the cosmic-ray muon and by the subsequent delta ray. 

\subsection{Electronics response}
In the LArPix system described in ref. \cite{Dwyer:2018phu} the pixel pads are uniquely instrumented by application-specific integrated circuits (ASICs), which provide a charge-sensitive amplifier (CSA) with self-triggered digitization. First, the signal from each pad is input to the CSA. Then, as signal accumulates on the pad, the voltage at the output of the CSA grows until it exceeds the discriminator threshold, which in our simulation is set to 28~mV, equivalent to the signal induced by $7\times10^3$ electron charges. The discriminator, in turn, triggers the output digitization through an 8-bit analog-to-digital converter (ADC). The ADC stores the CSA output voltage and then the CSA is reset, discarding the collected charge. The CSA is then ready to collect the subsequent signal. The minimum time between two consecutive ADC counts is 11 clock cycles, which with our 10 MHz clock correspond to \SI{1.1}{\micro\second}. The time between the threshold crossing and the signal digitization is tunable in the LArPix ASIC. In practice, this is 1.6~\si{\micro\second}, which is tuned to the typical signal size. While leakage current is negligible, sub-threshold charge could collect on the channel introducing a bias in the charge measurement. A periodic reset limits the impact of this effect.

Three kinds of noise are included in the simulation: (1) a reset noise, which corresponds to a random pedestal shift after every trigger reset, (2) a discriminator noise which affects the discriminator threshold, and (3) an uncorrelated noise. These are set in the simulation 900, 650, and 500 electron charges, respectively. These values have been tuned from the data described in section \ref{sec:data}.

\begin{figure}[htbp]
\centering
\includegraphics[width=0.7\textwidth]{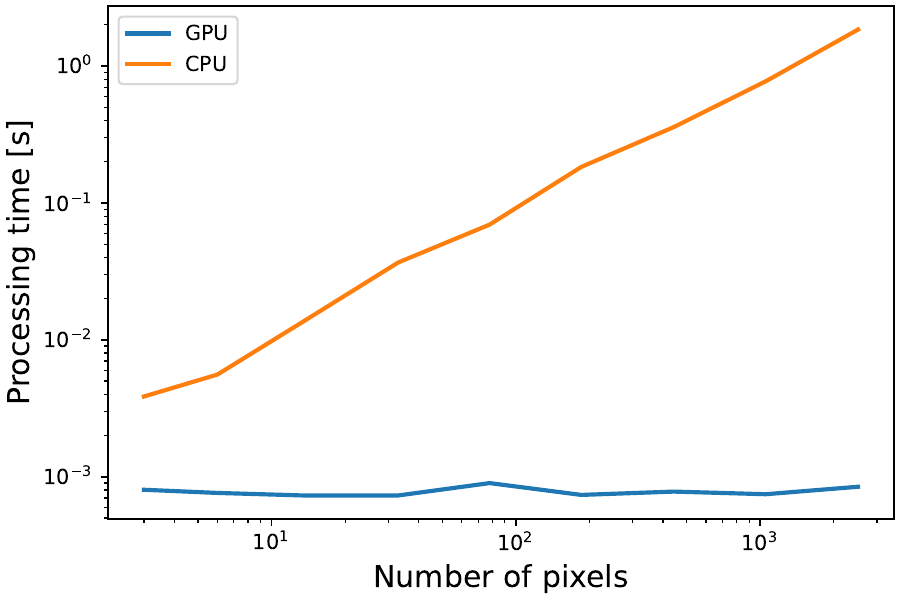}
\caption{Processing time for the simulation of the electronics response with the GPU implementation using a CUDA kernel (blue) and with a CPU implementation compiled with Numba with explicit parallelization of the loops (orange). For reference, a neutrino beam spill will induce a signal on around $5\times10^4$ pixels of the ND-LAr. The simulation has been performed on a node of the NERSC Cori supercomputer, which contains two sockets of 20-core Intel Xeon Gold 6148 (Skylake) at 2.40 GHz and 8 NVIDIA\textregistered~Tesla\textregistered~V100 (Volta) GPUs.}\label{fig:feecomparison}
\end{figure}

In our implementation, after summing the current induced by different tracks on the same pixel $p_i$, a CUDA kernel $K_{\mathrm{electronics}}(p_i)$ simulates the trigger logic and the noise sources described above. The threads are organized in a one-dimensional grid, where each thread processes in parallel the current induced on a single pixel $p_i$. The CUDA kernel is exponentially faster than the CPU implementation and is more than three orders of magnitude faster with $\mathcal{O}(10^3)$ pixels, as shown in figure \ref{fig:feecomparison}.

\begin{figure}[htbp]
\centering
\includegraphics[width=0.7\textwidth]{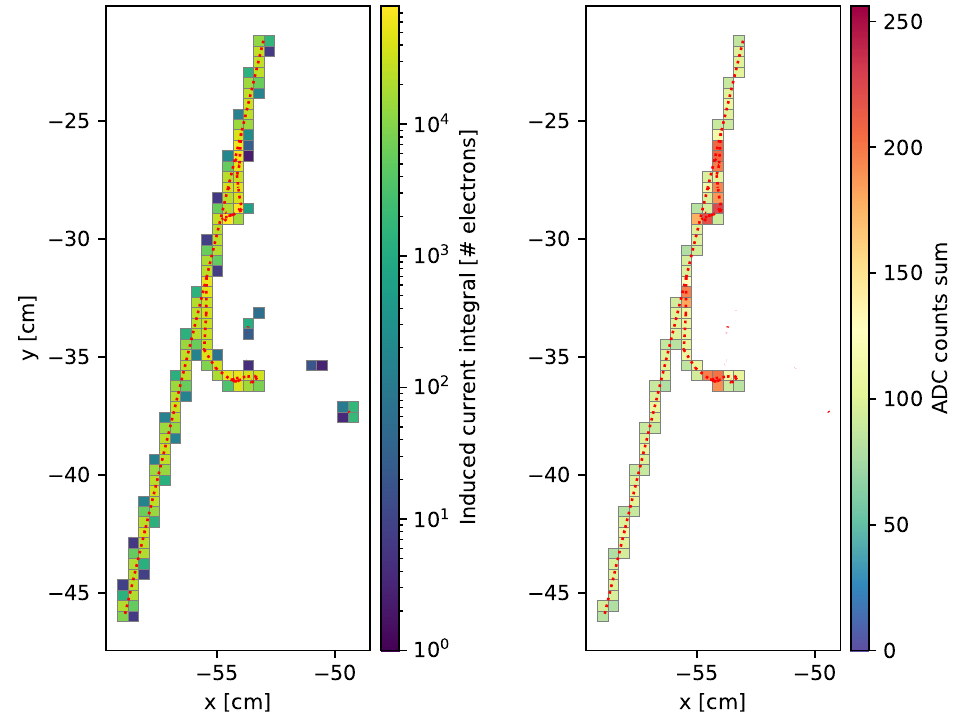}
\caption{Bi-dimensional projection of a cosmic-ray muon interacting in the detector with the integral of the current induced on each pixel (left) and the sum of the ADC counts after the digitization stage (right).}\label{fig:adc}
\end{figure}

Figure \ref{fig:adc} shows a comparison of the integral of the induced current on each pixel and the corresponding ADC counts for a cosmic-ray muon. When the integral signal on a certain pixel does not reach the discrimination threshold no ADC count is stored, so it is possible to have pixels with a non-null induced current but no ADC count.

The resulting ADC counts and their corresponding timestamps are stored in a bi-dimensional array, where one dimension is the pixel index and the other is the ADC count index. They are then saved in the same HDF5 format used by the LArPix readout system, facilitating comparison between data and simulation. 

\section{Light simulation}\label{sec:light}

\subsection{Incident light calculation}\label{sec:inc_light}
The scintillation light produced by the excitation of the liquid argon provides additional calorimetric and topological information and a fast timing signal. Within \texttt{larnd-sim}, we include GPU algorithms to model scintillation light production, timing, and the associated electronics response.

Light intensity is calculated using a self-consistent recombination model. Namely, each track segment $dE/dx$ is converted to a total number of emitted photons $N_\gamma$ using the same recombination parameterization used for the $dQ/dx$:
\begin{equation}
    \frac{dN_\gamma}{dx} = \left(W_\textrm{ph}^{-1} - \mathscr{R} W_\textrm{ion}^{-1}\right) \frac{dE}{dx}
\end{equation}
where $W_\textrm{ph}$ is the photon production per deposited energy at zero field, and $\mathscr{R}$ is from one of the recombination models described in section \ref{sec:drifting}.

Light propagation and material quantum efficiencies are pre-tabulated using a dedicated \textsc{Geant4} simulation into a LUT containing an average visibility $V_i\left[i_x,i_y,i_z\right]$ and relative time distribution $f_i\left[i_x,i_y,i_z,i_t\right]$ for each 3D voxel $\left(i_x, i_y, i_z\right)$ and each photosensor $i$ in the active volume~\cite{koller:2021}. For each track segment and photosensor, the total number of photoelectrons (p.e.) is calculated as
\begin{equation}
    N_{\textrm{pe}, ij} = \epsilon_i V_i\left[i_x,i_y,i_z\right] \left( \frac{dN_\gamma}{dx} \right)_j dx_j
\end{equation}
where $\epsilon_i$ is the overall detection efficiency of photosensor $i$ and $dx_j$ is the length of the track segment $j$. To implement this, we use a two-dimensional grid across the track segment index and detector index, such that each thread calculates the number of photoelectrons observed on a light detector from its respective track segment. In testing, we observe a speed-up factor of about $56\times$ by using the CUDA kernel for this task.

\subsection{Photocurrent simulation}\label{sec:photocurrent}

\begin{figure}[htbp]
     \centering
     \begin{subfigure}[b]{0.49\textwidth}
         \centering
         \includegraphics[width=\textwidth]{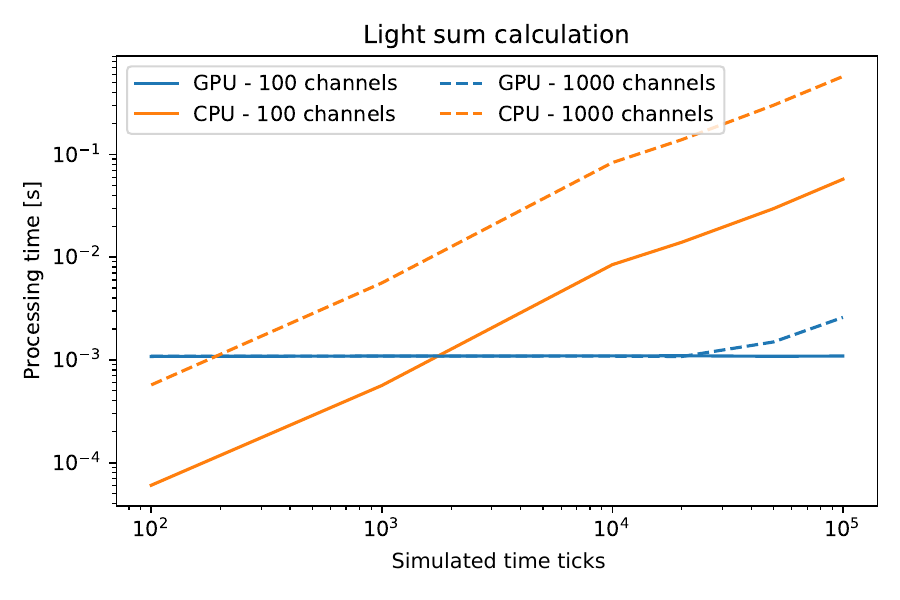}
         \caption{Comparison of light time profile calculation}
         \label{fig:light_sum_profile}
     \end{subfigure}
     \begin{subfigure}[b]{0.49\textwidth}
         \centering
         \includegraphics[width=\textwidth]{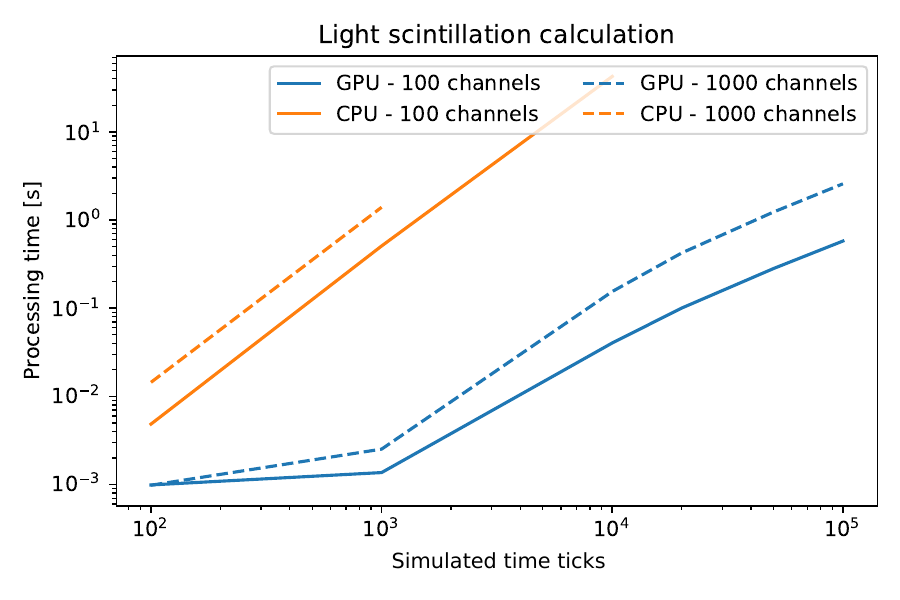}
         \caption{Comparison of scintillation profile convolution}
         \label{fig:light_scint_profile}
     \end{subfigure}
     \begin{subfigure}[b]{0.49\textwidth}
         \centering
         \includegraphics[width=\textwidth]{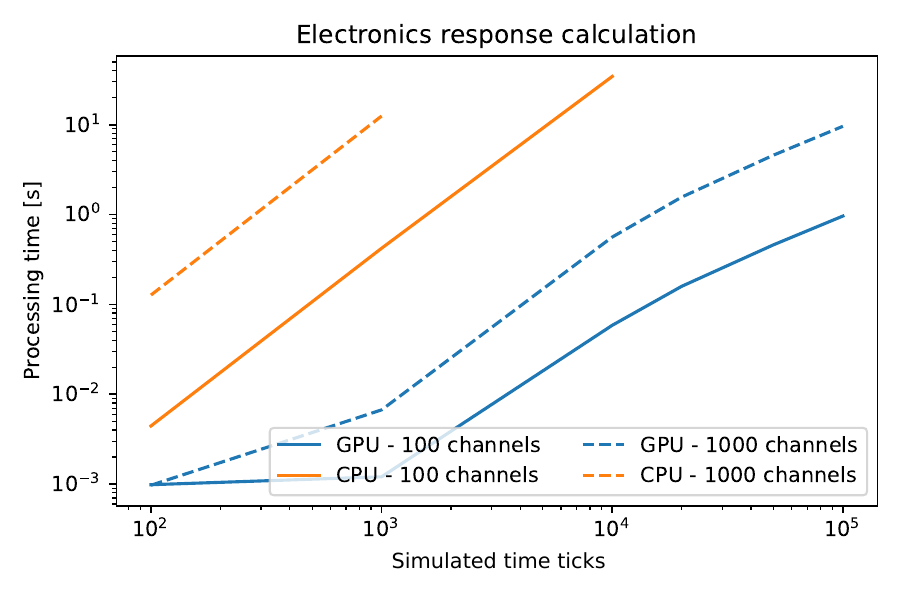}
         \caption{Comparison of electronics response convolution}
         \label{fig:light_response_profile}
     \end{subfigure}
     \caption{A summary of the runtime of the CUDA-compiled light simulation kernels described in the text using 100 identical track segments in the center of a TPC. The three most computationally intensive steps of the light simulation are compared to an equivalent parallelized but CPU-based calculation. For the light scintillation calculation and the electronics response calculation, the run time does not increase with more track segments. For reference, the drift time window of ND-LAr will be around $3\times10^3$ time ticks.}
     \label{fig:light_profile}
\end{figure}

Once the total light signal is determined for each track segment, they are summed into a photocurrent time profile on each photosensor using the time distribution stored within the LUT:
\begin{equation}
    I_{\textrm{pe},i}\left[i_t\right] = \frac{\Delta t_\textrm{LUT}}{\delta t} \sum_j f_i\left[i_x,i_y,i_z,i_t - t_0/\Delta t_\textrm{LUT}\right] N_{\textrm{pe}, ij},
\end{equation}
where $t_0$ is the segment deposition time, $\delta t$ is the simulation time tick, and $\Delta t_\textrm{LUT}$ is the time bin size of the LUT. The implementation of this calculation first determines the total number of ticks needed to fully simulate the event, with a configurable time pre- and post-buffer, and then pre-allocates the photocurrent time profile $I_{\textrm{pe},i}[i_i]$ array on the GPU. A CUDA kernel then iterates over the incident light array $N_{\textrm{pe}, i}$ summing the contribution at each time tick. To maintain independence between threads, we use a two-dimensional grid on the dimensions of the output time profile array. Each thread is responsible for calculating the sum over all track segments for a single photodetector and time tick. We observe a factor of 45 improvement by using the GPU kernel for this calculation at scale. However below approximately 1000 channels $\times$ 1000 ticks the parallel CPU calculation becomes faster due to the copying of static configuration data to the GPU, see figure~\ref{fig:light_sum_profile}.

After the time profile has been created, the simulation calculates the smearing effect due to the scintillation light emission. LAr scintillation light consists of two components: a prompt component with a characteristic decay time of~$\sim$7~ns; and a slow component produced with a characteristic decay {{time}} of~$\sim$1600~ns~\cite{kubota:1978,morikawa:1989}. These values, which are assumed to be constant in our simulation, depend on the concentration of non-argon impurities in the liquid~\cite{acciarri:2010a,acciarri:2010b} and on the electric field \cite{Segreto:2020qks}. A two-component exponential model for the scintillation time profile
\begin{align}
    f_\textrm{scint}(t) = \frac{f_p}{\tau_p} e^{-t/\tau_p} + \frac{1-f_p}{\tau_s} e^{-t/\tau_s}
\end{align}
is used to simulate the broadening of the light signals due to these processes, where $f_p$ is the fraction of total light arising from the prompt component ($\sim$0.3~\cite{hitachi:1983}), and $\tau_p$ and $\tau_s$ are the prompt and slow decay times, respectively.

The convolution is direct and truncated:
\begin{align}
    I_{\textrm{scint},i}[i_t] = \sum_{j_{t}=j_{t,\textrm{min}}}^{i_t} I_{\textrm{pe},i}[j_t] * f_\textrm{scint}( (i_t-j_t)\delta t),
\end{align}
where $j_{t,\textrm{min}}=i_t - 5\frac{\tau_t}{\delta t}$. The convolution kernel is truncated at $5\times$ the scintillation slow component lifetime. Within this interval, >99\% of the total light from an energy deposit will have been emitted, thus minimally impacting the accuracy of the simulation. However, by using a truncated convolution, we improve the scaling of the simulation from $\mathcal{O}(N^2)$ to $\mathcal{O}(N\times N_\textrm{trunc})$, where $N_\textrm{trunc}$ is the number of truncated time ticks. This convolution is implemented in a CUDA kernel and threads are distributed in a two-dimensional grid across the light detector index $i$ and the time tick index $i_t$, such that each thread calculates the weighted sum across the input array $I_{\textrm{pe},i}$. The direct convolution of the scintillation light profile along with the electronics response convolution, discussed later, are the most computationally intense stages of the light simulation. Figure~\ref{fig:light_scint_profile} and figure~\ref{fig:light_response_profile} show the performance of these kernels with the number of time ticks simulated and the number of light channels. Within the event sizes tested, regardless of the number of time ticks and light channels, a significant speed-up is realized by moving this calculation to the GPU.

Fluctuations in the number of photoelectrons produced by the incident light due to counting statistics is included for each time profile bin. An inverse transform sampling Poisson random number algorithm~\cite{devroye:1986} is used to generate the number of observed photoelectrons per time tick. The average number of steps required by this algorithm scales in proportion to the mean of the distribution. To improve the runtime for large signals, at greater than 30 expected photoelectrons, a normal distribution $\sim\mathcal{N}(I_{\textrm{scint},i}[i_t] \delta t,\sqrt{I_{\textrm{scint},i}[i_t] \delta t})$ is used. The choice of 30 photoelectrons is selected to limit the error from this approximation to less than 1\%.

\subsection{Electronics response}\label{sec:light_response}

\begin{figure}[htbp]
     \centering
     \begin{subfigure}[b]{0.7\textwidth}
         \centering
         \includegraphics[trim={0 90 0 90}, width=\textwidth]{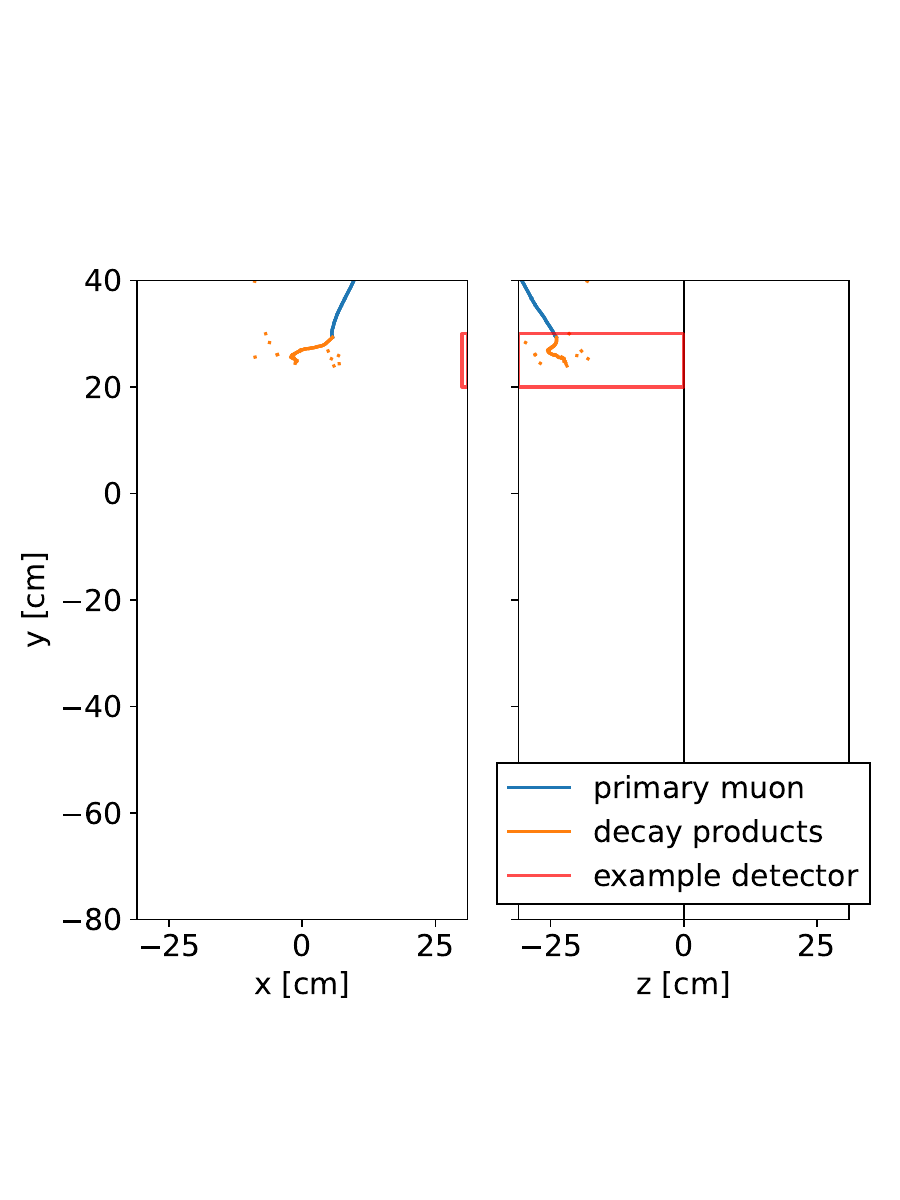}
         \caption{2D projection of a simulated stopping muon and subsequent decay.}
         \label{fig:light_example_2d}
     \end{subfigure}
     \\
     \begin{subfigure}[b]{0.45\textwidth}
         \centering
         \includegraphics[width=\textwidth]{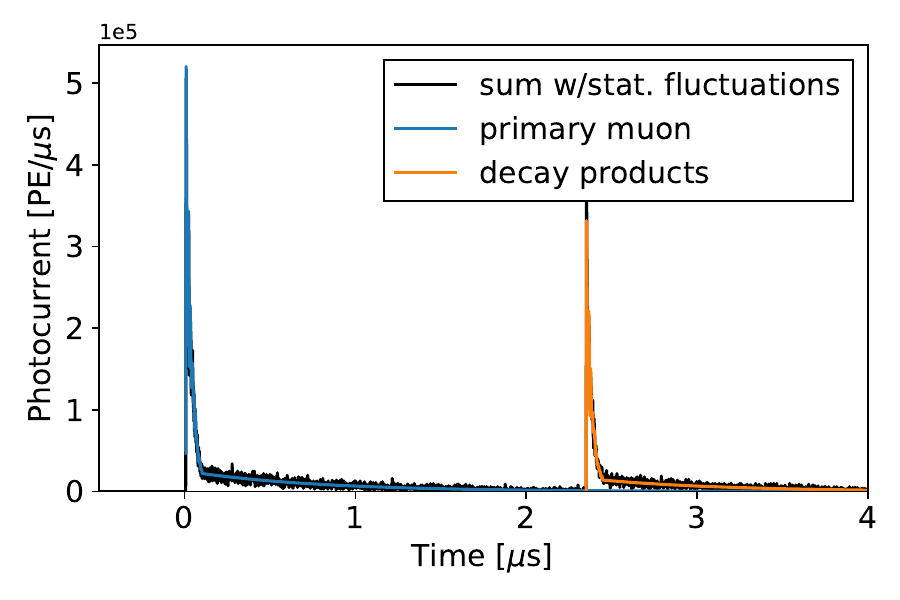}
         \caption{Photocurrent on example photosensor after including LUT and scintillation time smearing.}
         \label{fig:light_example_scintillation}
     \end{subfigure}
     \hspace{2.5em}
     \begin{subfigure}[b]{0.45\textwidth}
         \centering
         \includegraphics[width=\textwidth]{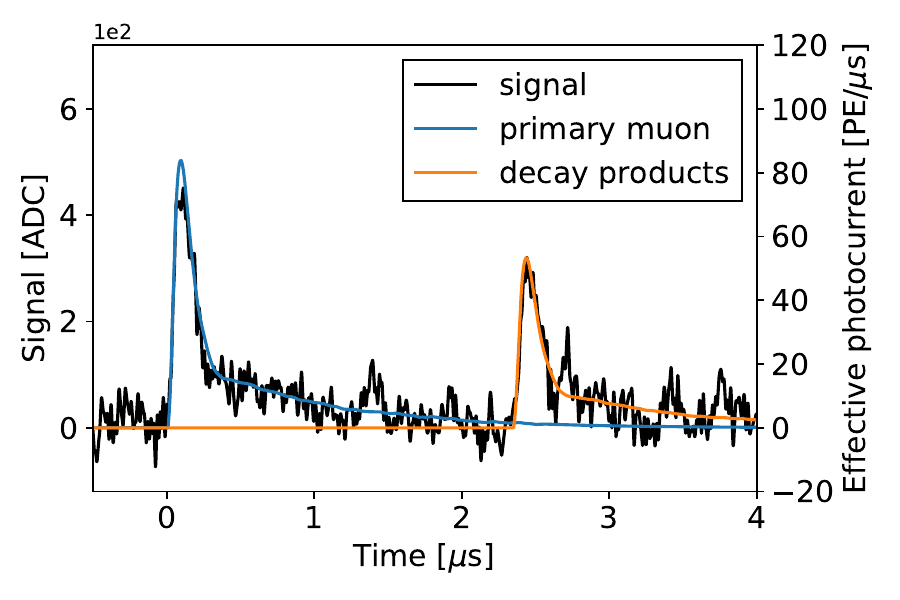}
         \caption{Photosensor output and effective photocurrent after applying response and noise models.}
         \label{fig:light_example_response}
     \end{subfigure}
     \caption{Example simulated muon decay and subsequent light simulation sequence. The region highlighted in figure~\ref{fig:light_example_2d} corresponds to the location of the light detector simulated in figure~\ref{fig:light_example_scintillation} and \ref{fig:light_example_response}.}
     \label{fig:light_example}
\end{figure}

The electronics response of the light detector readout is simulated assuming a perfectly linear ADC, a user-specified unit-normalized impulse response model, and purely uncorrelated noise:
\begin{align}
    V_i[i_t] = n_i[i_t] + \sum_{j_t=0}^{N_\textrm{resp}} G_i I_{\textrm{scint},i}[j_t] i_\textrm{resp}[i_t - j_t] \frac{\Delta t_\textrm{resp}}{\delta t}
\end{align}
where $n_i[i_t]$ is the simulated noise, $i_\textrm{resp}[i_t]$ is the unit-normalized response model, $N_\textrm{resp}$ is the number of bins in the response model, and $G_i$ is the individual photosensor gain with units of [ADC $\cdot$ time / PE]. Noise is generated individually on each light detector by using a user-specified fast-fourier transform of the noise distribution to generate random phase sinusoids with appropriate amplitudes. Figure~\ref{fig:light_example} demonstrates the simulation sequence on an example muon decay event with non-trivial time structure.

Each time profile is segmented into individual light triggers using a direct threshold on the sum of multiple light detectors. A serial algorithm loops over each light detector group and identifies each threshold crossing, accounting for dead-time between triggers. Because the timing of later triggers depends on the timing of earlier triggers, this algorithm cannot be parallelized. However, the trigger simulation only contributes $\sim3$\% to the overall simulation runtime and scales with the number of time ticks simulated, thus the use of a serial algorithm for the trigger generation is not a significant contributor to the overall simulation time at the DUNE Near Detector regime of <100 triggers per spill per light detector group.

The final ADC waveform is then linearly interpolated into the desired sampling frequency and ADC resolution. Trigger information and waveforms are saved into dedicated datasets within the same output HDF5 file as the charge simulation.

\subsection{Truth propagation}\label{sec:light_truth}

True particle energy deposition information can be propagated through this chain to provide the true track segment id and incident photoelectrons contributing to each ADC sample. However for typical light detectors with $\mathcal{O}(\si{\centi\meter})$ spatial resolution and relatively broad response times $\mathcal{O}(\si{\micro\second})$, most energy deposits originating from an interaction overlap. Thus, the number of truth entries needed scales as ($dN_\textrm{segments,typ}/dV$) $\times$ ($N_\textrm{samples}$) $\times$ ($N_\textrm{photosensors}$) $\times$ (visible volume). Using typical numbers of $\mathcal{O}(500/\mathrm{event~m}^3)$, $\mathcal{O}(100)$, $\mathcal{O}(100)$, and $\mathcal{O}(10~\mathrm{m}^3)$, respectively, tracking the true track segment id and intensity at each sample necessitates $\mathcal{O}(5\times10^7)$ truth entries, using a minimum of $\mathcal{O}(400~\mathrm{MB/event})$ assuming each record consists of a 32-bit integer and 32-bit floating point number. Since the truth information for an event is highly correlated in time, we opt to by default only preserve the true number of photoelectrons and arrival time per track segment per light detector, reducing the required space by a factor of $N_\textrm{samples}$ and runtime by a factor of $N_\textrm{segments}/\textrm{tick}$. We however maintain the option to track the full truth information as needed by setting a configuration flag.

\section{Profiling}
The \texttt{larnd-sim} code has been profiled using the NVIDIA Nsight\textsuperscript{\texttrademark} Systems and the NVIDIA Nsight\textsuperscript{\texttrademark} Compute tools v2021.3.0 with an input dataset of 250 simulated cosmic rays. The analysis shows that the algorithm is well optimized for GPUs. The charge simulation is responsible for around 70\% of the processing time. Of this 70\%, 97.5\% of the time is used by the induced current CUDA kernel described in section \ref{sec:induced}. This kernel, in turn, spends 99.9\% of the processing time for computing operations. The time spent for memory transfers, including also the initial time to allocate the dataset on the GPU memory, can be considered negligible. 

\begin{figure}[htbp]
\centering
\includegraphics[width=0.8\textwidth]{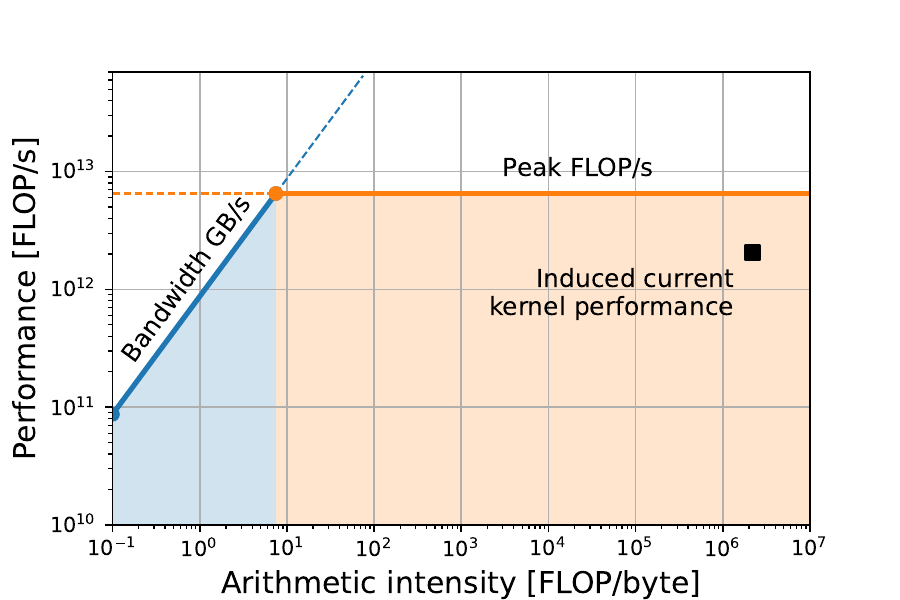}
\caption{Roofline plot for the induced current CUDA kernel (black square) on a GPU node of the NERSC Cori supercomputer, equipped with 8 NVIDIA\textregistered~Tesla\textregistered~V100 (Volta) GPUs. The kernel performances are far from the memory-bound region (in light blue) and close to the peak FLOP/s limit (in orange) of the GPU being used.}\label{fig:roofline}
\end{figure}

A Roofline \cite{roofline} analysis performed on the induced current CUDA kernel proves that the algorithm is compute-bound, so the time needed to complete its task is determined principally by the computing speed. Figure \ref{fig:roofline} shows that kernel performance for 10 cosmic-ray events with a total of 5013 active pixels, which is a typical amount of data processed for each GPU iteration, is close to the peak FLOP/s limit (in orange) and far from the memory bandwidth limit (in blue) of the GPU being used.

Overall, for a 0.5~m$^3$ LArTPC operated with a surface-level cosmic ray rate, the inclusion of the light simulation increases the runtime, the memory usage, and the overall data volume of the output by 30\%, 15\%, and 20\%, respectively.

\section{Simulation of a cosmic-ray sample and data comparison}\label{sec:data}
A single-phase LArTPC called the \emph{ArgonCube Module-0 Demonstrator} was operated in spring 2021 at the University of Bern as a prototype for the ND-LAr. The detector is divided into two functionally identical TPC drift regions, sharing a central high-voltage cathode that provides the drift electric field. Each anode is equipped with an array of $2\times4$ LArPix tiles of size $31.038\times31.038$~\si{cm^2}. A total of 4900 pixel pads of size $0.4434\times0.4434$~\si{cm^2} are etched on the side of each tile facing the active volume. Each tile is instrumented by 100 ASICs placed on the opposite side of the board. The LArTPC drift length is 30.27~\si{cm}. 

A dataset of cosmic-ray events was acquired with an applied electric field of $0.50$~\si{kV/cm}. The detector performances are analyzed in detail in ref. \cite{module0}.


Here we will compare the cosmic-ray data acquired with this experimental setup to a sample generated by the \texttt{larnd-sim} software. 
As a first step a dataset of cosmic rays is produced using the CORSIKA cosmic-ray generator v7.7410 \cite{Heck:1998vt}, using FLUKA 2011 for the low-energy hadronic interaction \cite{Battistoni:2015epi}. Then, the passage of the cosmic rays through the Module-0 geometry is simulated using \texttt{edep-sim}, a C++ wrapper around \textsc{Geant4}. The output of \texttt{edep-sim} is passed to \texttt{larnd-sim}, which produces a HDF5 sample in the same format of the Module-0 data.

\begin{figure}[htbp]
\centering
\includegraphics[width=0.95\textwidth]{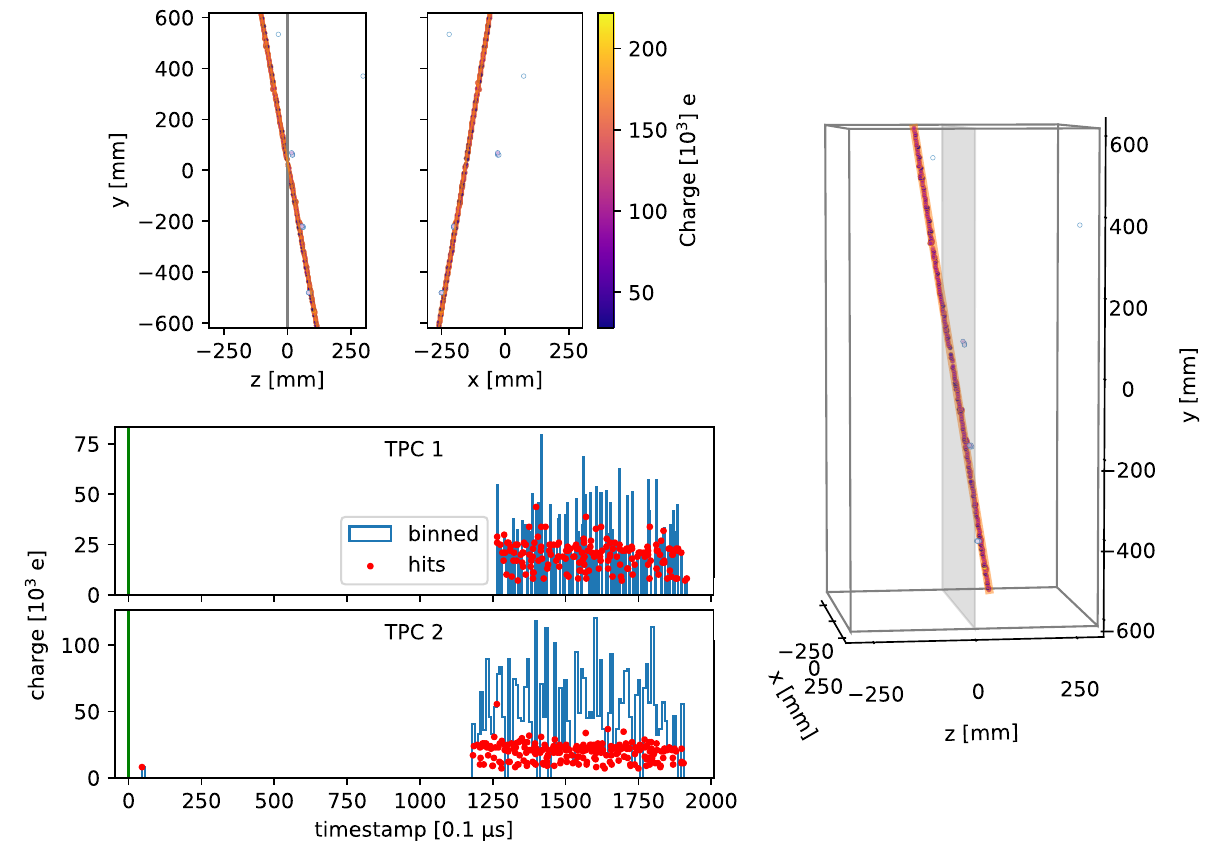}
\caption{The ADC counts are transformed into three-dimensional hits by a reconstruction script. The hits are grouped in clusters and a fed to a principal component analysis, which returns a reconstructed track. This plot shows, clockwise from top left, the hits for a simulated cosmic ray on the $zy$ and $xy$ planes, a three-dimensional event display, and the hits charge as a function of time.}\label{fig:evd}
\end{figure}

Both the simulation and the data samples are passed to a reconstruction script, which takes as input an array of three-dimensional hits, one for each ADC count. The $xy$ coordinates are given by the position of the pixel center on the tile, and the $z$ coordinate is given by the corresponding clock tick, using the inverse formula of eq. \eqref{eq:drift}. The hits are clustered together using the DBSCAN algorithm \cite{dbscan}. Hits grouped in the same cluster are fed to a principal component analysis (PCA), which returns three-dimensional segments that we define as \emph{reconstructed tracks}. Figure \ref{fig:evd} shows an example of a simulated cosmic ray crossing the cathode plane after the reconstruction stage, with the reconstructed track and the corresponding hits. 

To improve the system self-trigger stability, 7.8\% of the pixels in Module-0 were disabled due to an understood grounding issue, mostly near the tiles edges. The same pixels were also disabled in the simulation. Thus, reconstructed tracks may show artificial gaps due to the presence of disabled channels. Also, cathode-piercing tracks can be reconstructed as separated tracks, due to the non-null cathode thickness. 
In order to reconstruct the original cosmic-ray trajectories, 
we iteratively stitch together pairs of reconstructed tracks, {{based on their relative directions and positions at the cathode. To set the threshold for these metrics, a subsample of data events are reconstructed a second time after randomly translating and masking hits that fall into the disabled regions. We require that 95\% of the gaps produced in this procedure are correctly identified and resolved.}}


Module-0 was operated principally in two channel threshold regimes, low and high, which correspond to thresholds of around 6000 and 12000~$e^-$, respectively. Here we compare the low-threshold data to the simulated sample, which was produced with a discrimination threshold roughly equivalent to the low regime one. 



\begin{figure}[htbp]
\centering
\includegraphics[width=0.7\textwidth]{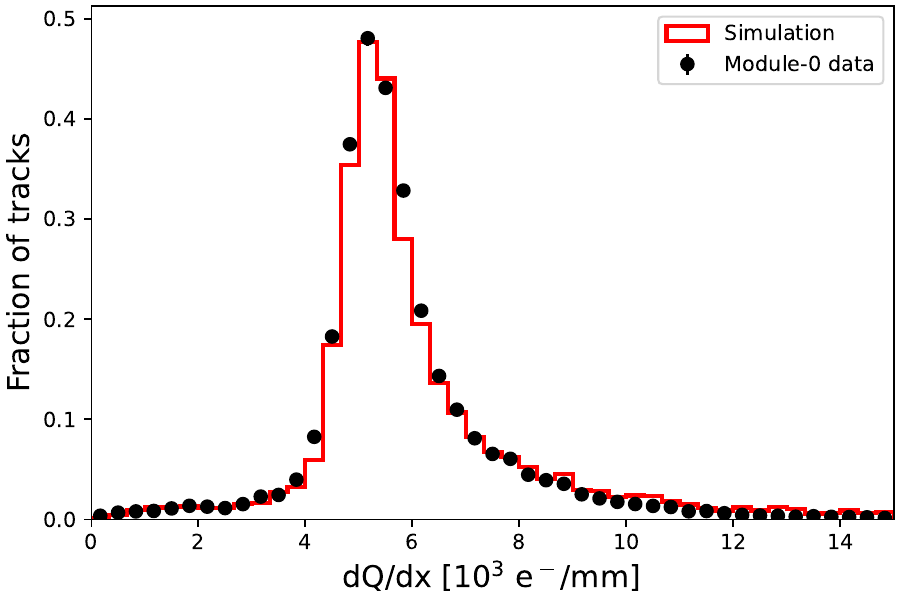}
\caption{$dQ/dx$ measured in data and simulation for reconstructed tracks with at least 10 associated hits. The distributions have been normalized to unity.}\label{fig:dqdx_fit}
\end{figure}

The amount of charge deposited per unit length ($dQ/dx$) is measured both in data and simulation using the reconstructed tracks, which are required to have at least 10 associated hits. In the simulation, the signal amplitude of the CSA is calibrated using a fixed gain of 250~e$^{-}$/\si{\milli\volt}. The value of the gain in the data is obtained by performing a template fit of the $dQ/dx$ distribution with the simulated one, shown in figure \ref{fig:dqdx_fit}. Here, the $dQ$ is the sum of the hits charge associated to the reconstructed track and $dx$ is the length of the reconstructed track. The best-fit value of 219~e$^{-}$/\si{\milli\volt} is compatible with the value of 221~e$^{-}$/\si{\milli\volt}, obtained with a dedicated stand-alone test of the LArPix ASIC CSA.

Reconstructed tracks can be sub-divided into segments of variable length, from 10 to 400~mm. The deposited charge $dQ$ is the sum of the hit charges belonging to the same segment and $dx$ is the length of the segment. The distributions have been fitted with a Gaussian-convoluted Moyal function \cite{Moyal:2009xna}.

\begin{figure}[htbp]
\centering
\includegraphics[width=0.96\textwidth]{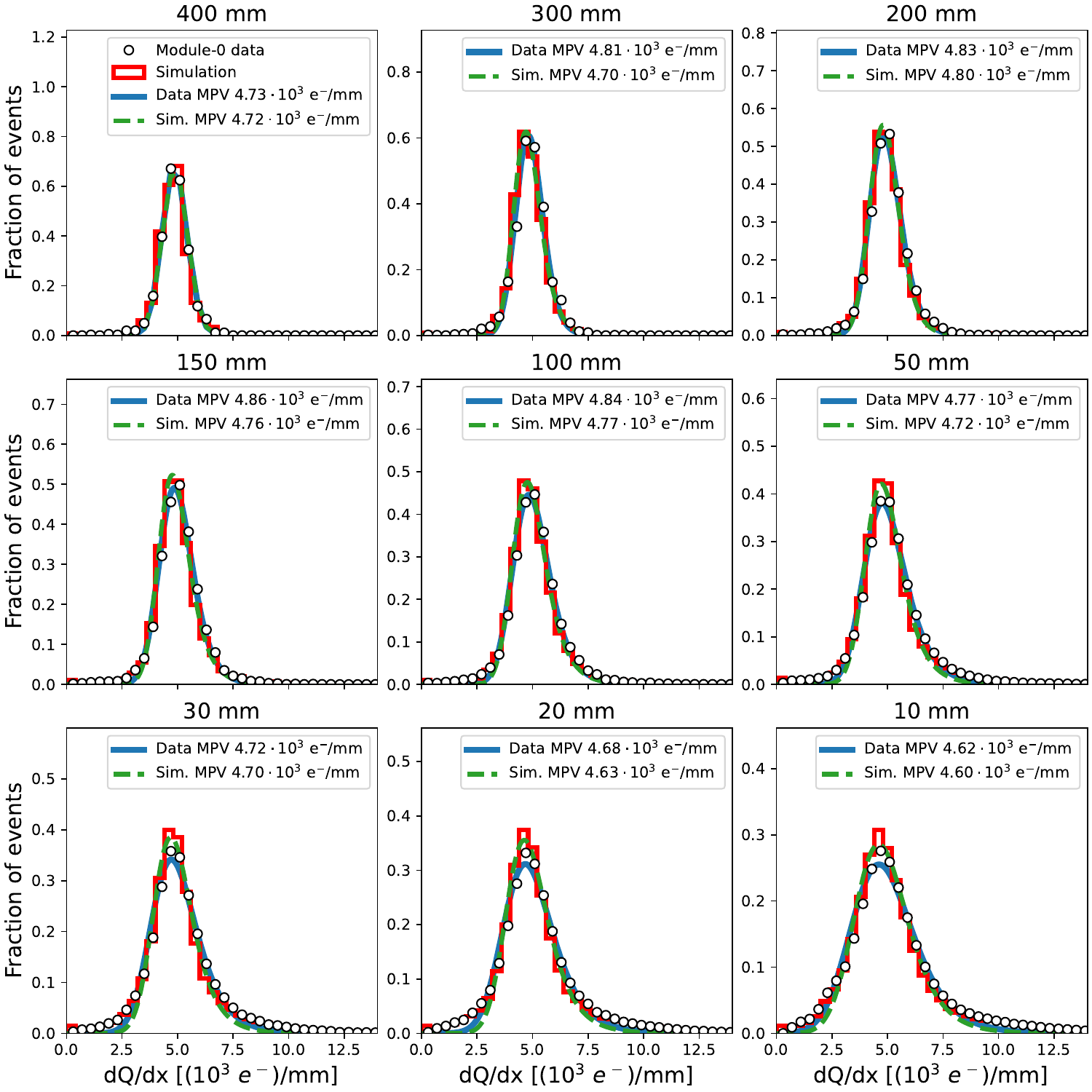}
\caption{$dQ/dx$ measured for segments of different lengths for low threshold runs (white dots) and a sample of simulated cosmic rays (red line). The distributions have been normalized to unity and fitted with a Gaussian-convoluted Moyal function for data (solid blue line) and simulation (dashed green line). The legend shows the most probable value (MPV) for each fit.}\label{fig:dqdx}
\end{figure}

Figure \ref{fig:dqdx} shows the $dQ/dx$ data and simulated distributions, as well as their respective fits. 
The $dQ/dx$ distributions, normalized to the number of tracks, are in good agreement with data and simulation for all segment lengths.


\section{Conclusions}
We show that it is possible to implement the simulation of a pixelated LArTPC using highly-parallelized GPU algorithms. The algorithms are written in Python and they are compiled on the GPU using the Numba just-in-time compiler. 

The software provides an end-to-end microphysical simulation for light readout and pixelated charge readout.
Table \ref{tab:recap} shows a summary of the speed-up achieved by the GPU implementation for various steps of the simulation. 

\begin{table}[htbp]
\centering
\caption{\label{tab:recap} Summary of speed-up factors achieved by the GPU implementation for various simulation steps.}
\smallskip
\begin{tabular}{|llr|r|}
\hline
Calculation & Loop over & Quantity & GPU speed-up factor\\
\hline
Recombination factor & Segments & $10^5$ & $\times3$\\
Induced current & Pixels & $10^3$ & $\times7314$\\
Charge electronics response & Pixels & $10^3$ & $\times985$\\
Light time profile & Time ticks & $10^5$ & $\times228$\\
Scintillation profile & Time ticks & $10^3$ & $\times568$\\
Light electronics response & Time ticks & $10^3$ & $\times1883$\\
\hline
\end{tabular}
\end{table}

In general, trivial algorithms (such as the calculation of the recombination factor) require a large amount of input data before the GPU implementation starts being faster. This is expected, since there are launch and execution overheads associated with CUDA kernels. For more complex algorithms, such as the simulation of the charge readout, the GPU generally starts being may orders of magnitude faster than the CPU with only a few tens of simulated pixels in the detector. In particular, the calculation of the induced current, which is the most computationally expensive task, is around four orders of magnitude faster for $\mathcal{O}(10^3)$ pixels. Likewise, the simulation of convolution-heavy light signals is faster on all event sizes tested, typically achieving three orders of magnitude improvement. With the acceleration achieved, we are able to simulate the charge (light) signals on $O(10^6)$ ($O(10^3)$) individual channels with sub-microsecond (nanosecond) resolution and in timescales reasonable for large Monte Carlo simulations.

As an example, the full simulation of $10^4$ cosmic rays in a ton-scale LArTPC takes in total approximately 450~s, 100~s for the simulation of the passage of the initial particles through matter with \texttt{edep-sim}, and the remaining 350~s for the detector simulation with \texttt{larnd-sim}. Highly-parallelized simulation algorithms such as the one described in this document could be adapted to speed up the simulation of the DUNE Far Detector as well, which will have $\mathcal{O}(10^6)$ readout channels \cite{DUNE:2020txw}. 

We have also performed the comparison of a simulated sample of cosmic rays with the data acquired by a prototype LArTPC equipped with a large-scale pixelated readout system, called the ArgonCube Module-0 Demonstrator. Charge distributions in data and simulation are in general good agreement. 

The Module-0 LArTPC was moved from Bern to Fermilab in October 2021, where it will be tested on the NuMI neutrino beamline with three other identical modules, currently under construction \cite{module0}, {{forming the ND-LAr 2x2 prototype}}. The production of a dataset of accurately simulated neutrino interactions will be of fundamental importance for the analysis of this measurement. \texttt{larnd-sim} will also be used to simulate the detector response of the full ND-LAr detector and has been included into the DUNE Offline Computing Conceptual Design Report \cite{DUNE:2022fcw}. {The DUNE collaboration is already using \texttt{larnd-sim} on the NERSC Perlmutter supercomputer, which shares a similar GPU node architecture with Cori, to perform ND-LAr 2x2 prototype simulations.} A workflow chaining together \texttt{edep-sim}, \texttt{larnd-sim} and a machine learning-based reconstruction stage is currently under development.


The choice of a high-level and largely supported programming language such as Python can enable the implementation of a fully differentiable simulator in the near future, by exploiting libraries commonly used in the machine learning and artificial intelligence fields (TensorFlow \cite{tensorflow2015-whitepaper}, JAX \cite{jax2018github}). A differentiable model will be able to use a gradient-based optimization, such as gradient descent, to automatically infer the detector simulation input and the detector physics model parameters. This technique has already been actively explored for the simulation of the photon propagation in a LArTPC \cite{siren}.

\acknowledgments


%
%
This document was prepared by the DUNE collaboration using the
resources of the Fermi National Accelerator Laboratory 
(Fermilab), a U.S. Department of Energy, Office of Science, 
HEP User Facility. Fermilab is managed by Fermi Research Alliance, 
LLC (FRA), acting under Contract No. DE-AC02-07CH11359.

%
%
This work was supported by
CNPq,
FAPERJ,
FAPEG and 
FAPESP,                         Brazil;
CFI, 
IPP and 
NSERC,                          Canada;
CERN;
M\v{S}MT,                       Czech Republic;
ERDF, 
H2020-EU and 
MSCA,                           European Union;
CNRS/IN2P3 and
CEA,                            France;
INFN,                           Italy;
FCT,                            Portugal;
NRF,                            South Korea;
CAM, 
Fundaci\'{o}n ``La Caixa'',
Junta de Andaluc\'ia-FEDER,
MICINN, and
Xunta de Galicia,               Spain;
SERI and 
SNSF,                           Switzerland;
T\"UB\.ITAK,                    Turkey;
The Royal Society and 
UKRI/STFC,                      United Kingdom;
DOE and 
NSF,                            United States of America.
%
%
This research used resources of the 
National Energy Research Scientific Computing Center (NERSC), 
a U.S. Department of Energy Office of Science User Facility 
operated under Contract No. DE-AC02-05CH11231.

We are also thankful to Max Katz from NVIDIA\textregistered~and Daniel Margala from NERSC for the precious suggestions.

\bibliographystyle{JHEP}
\bibliography{biblio}{}

\end{document}